\documentclass[onecolumn,traditabstract]{aa}

\usepackage{natbib}
\usepackage[english]{babel}
\usepackage{algorithmic}
\usepackage{algorithm}
\usepackage{theorem}
\usepackage{amssymb}
\usepackage{amsmath}
\usepackage{mathtools}
\usepackage{graphics}
\usepackage{epsfig}
\usepackage{subfigure}
\usepackage{fullpage}
\usepackage{url}
\usepackage{multirow}
\usepackage{color}

\newcommand{\minimize}{\mathop{\operatorname{minimize}}\limits}
\newcommand{\tb} {\textbf}
\DeclareMathAlphabet\mbi{OML}{cmm}{b}{it}
\DeclareMathAlphabet\mi{OML}{cmm}{sb}{it}

\newcommand{\mb} {\mathbf}

\newcommand{\Rn}[1]{\ensuremath{\mathbb{R}^{#1}}}

\newcommand{\fracd}[2]{\ensuremath{\displaystyle\frac{#1}{#2}}}

\makeatletter

\renewcommand{\lim}{\mathop{\operatorname{lim}}\limits}

\usepackage{amstext,amssymb,amsbsy,amsmath,amscd,mathrsfs,dsfont}

\begin{document}

\title{Sparse point-source removal for full-sky CMB experiments: application to WMAP 9-year data}

\author{F.C. Sureau, J.-L. Starck, J. Bobin, P. Paykari, A. Rassat}

\institute{
  Laboratoire AIM, UMR CEA-CNRS-Paris 7, Irfu, SAp/SEDI, Service d'Astrophysique, CEA Saclay, F-91191 GIF-SUR-YVETTE Cedex, France}

\date\today

\abstract{Missions such as \emph{WMAP} or \emph{Planck} measure full-sky fluctuations of the cosmic microwave background and foregrounds, among which bright compact source emissions cover a significant fraction of the sky. To accurately estimate the diffuse components, the point-source emissions need to be separated from the data, which requires a dedicated processing. We propose a new technique to estimate the flux of the brightest point sources using a morphological separation approach: point sources with known support and shape are separated from diffuse emissions that are assumed to be sparse in the spherical harmonic domain. This approach is compared on both \emph{WMAP} simulations and data with the standard local $\chi^2$ minimization, modelling the background as a low-order polynomial. The proposed approach generally leads to 1) lower biases in flux recovery, 2) an improved root mean-square error of up to $35\%,$ and 3) more robustness to background fluctuations at the scale of the source. The \emph{WMAP} 9-year point-source-subtracted maps are available online. } 

\keywords{Cosmology : CMB, Data Analysis, Methods : Statistical}
\maketitle

\section{Introduction}

Missions such as \emph{Planck} and \emph{WMAP} provide high-resolution full-sky maps of microwave emissions. The primary goal of these missions is to measure the fluctuations in the cosmic microwave background (CMB), the relic radiation of the big bang, and hence provide key information about the birth and evolution of our Universe. These missions also provide full-sky information on the Galactic and extra-Galactic emissions in a frequency range that has not been probed before. One of the main scientific challenges in the CMB data analysis consists of accurately separating the CMB, the various Galactic emissions (synchrotron, free-free, dust and Galactic compact-source emissions, to name a few) as well as the contribution from extra-Galactic sources. In particular, bright compact-source emissions cover a significant fraction of the sky \citep{ERCSC11,Wright09}, even at high Galactic latitudes, where the CMB is less affected by other diffuse foregrounds. These emissions therefore need to be carefully dealt with for CMB or Galactic data analysis.

From a source-separation perspective, these emissions are difficult to model since the compact sources display both spectral and temporal variability. As reported in \citet{Wright09}, at least one-third of the extra-Galactic \emph{WMAP} sources display temporal variability with high confidence, with more than a 2:1 range in fluxes. On the other hand, extrapolating the fluxes from catalogues obtained at lower frequency than that probed by \emph{WMAP} or \emph{Planck} is erroneous for radio sources \citep{PlanckXIV}: flat-spectrum emissions from the core of extra-Galactic sources are dominating for Planck or WMAP channels, whereas steep-spectrum lobe emissions dominate at lower frequencies. 
Finally, their high spectral variability makes it very difficult to estimate their flux with generic source-separation techniques that implicitly rely on the factorization of spatial and spectral information \citep{Tegmark98,Vielva03,Lopez06}. This means that
 each CMB data set needs to be processed independently to accurately estimate the compact-source fluxes in this data set.

The most distinctive information for separating compact and diffuse emissions is based on morphology, in particular for point sources where the shape of the emission is the point spread function (PSF) of the instrument at the wavelength considered. Consequently, specific pipelines were developed based on this information in the CMB data analysis; the compact sources are first detected and flux-estimated and then are either masked or their contribution subtracted before component separation is performed \citep{PlanckCompsep13}.

Numerous approaches based on the morphology of the sources have been proposed to address the initial problem of source detection in the CMB data: 
morphological image-processing such as Sextractor \citep{astro:bertin96}, matched filters \citep{Tegmark98,Barreiro03,Lopez06}, 
wavelet-based techniques \citep{gauss:cayon00, Vielva01, Martinez-Gonzalez03,Barreiro03,Lopez06}, and Bayesian detection \citep{Hobson03, Savage06, Carvalho07, Guglielmetti09, Argueso11}. 
Multichannel techniques, which take into account either an estimate of the power spectrum of diffuse emissions \citep{Herranz08} or the alleged spectral signatures for the point sources \citep{Lanz10, Vio12} have also been devised. Other multichannel techniques detect the point sources by cancelling the CMB contribution in the data by a weighted sum of the channels \citep{Wright09,Ramos11,Scodeller12}. We refer to \citet{CMBDataAnalysis} for a brief review of the techniques. Some compact-source catalogues have been published for the \emph{WMAP} \citep[for instance,][]{Wright09,ChenWright09,Ramos11,Scodeller12,WMAP9} or \emph{Planck} \citep{ERCSC11,ERCSC13} missions.

When bright sources are detected, their contribution can be removed from the data. Several strategies have been proposed.
\begin{itemize}
\item{Masking:} this is a standard approach where a point-source mask is applied to the data map before it is analysed. This solution is very efficient for CMB power spectrum estimation, and the price to pay is only a small loss of sensitivity at low multipoles \citep{Nolta09, Larson11}. However, for CMB non-Gaussianity analysis the mask may create spurious features in the non-Gaussianity estimators. In addition, masking complicates the Galactic diffuse emission analysis
\item{Masking and inpainting:} to avoid artefacts caused by the mask on large scales,  the masked data can be
interpolated using inpainting techniques. Several methods have been proposed, such as diffuse inpainting \citep{lfi2011}, CMB-constrained realization inpainting \citep{bucher2012,kim2012}, or sparse inpainting \citep{inpainting:abrial06,starck:abrial08,starck2012}. It was shown that sparse inpainting, based on the sparsity of the CMB in spherical harmonics, does not significantly impact non-Gaussianity measures such as the skewness and kurtosis of the CMB data \citep{inpainting:abrial06} or the integrated Sachs-Wolfe signal \citep{starck:dupe2011} and the weak-lensing signal \citep{perotto10}. These inpainting approaches are, however, expected to be inaccurate in estimating the CMB in the masked regions, which would also be detrimental to Galactic studies.
\item{Fitting:} a more progressive approach is to first model the background due to diffuse emissions, then model the shape of the compact sources by modelling the instrument PSF, and finally estimate the flux by aperture photometry or PSF fitting. This is typically performed when the flux and the position are not jointly estimated in the detection algorithm, as performed in the Bayesian detection approaches (where the background can be modelled through a covariance matrix or described by a low-order polynomial).
This method was adopted by the \emph{Planck} and \emph{WMAP} teams to estimate the point-source flux \citep{ERCSC11,ERCSC13,Wright09,WMAP9}, 
where the background was modelled as a low-order polynomial (baseline or linear background). The few background parameters and the point-source flux were then estimated locally by $\chi^2$ minimization. 
\end{itemize}

\citet{Scodeller2012} showed that removing the point-source fluxes instead of masking them provides similar results on the final CMB power spectrum estimation with no biases on higher order statistics. A {\it cleaned} CMB map is much easier to analyse than a masked CMB map, since there is no need to handle the point-source mask anymore. Another advantage is that it provides a CMB estimate at the point-source positions in the mask (or a point-source-free channel estimate if applied on a channel instead of on the reconstructed CMB map), useful in particular for Galactic studies, while inpainting techniques only try to fill the gaps without destroying the statistical properties of the map.

In this paper, we propose a new approach for point-source removal that is based on a morphological source-separation method, assuming the sources have already been detected. We use a more flexible (and potentially more complex) model for the background to capture its fluctuations more accurately at the scale of the sources. In this approach we assume that the diffuse emissions are sparse in the spherical harmonic domain, while the point sources are sparse in the direct domain, and resolved compact sources are sparse in the wavelet domain. In section \ref{sec:methods}, we describe this method along with the proposed algorithm, which is adapted from recent convex-optimization techniques that solve the corresponding inverse problem. Results on compact-source removal in full-sky \emph{WMAP} realistic simulations are presented in section \ref{sec:results}, where our approach is compared with the standard flux-fitting or low-order polynomial background-fitting as performed by \emph{WMAP} and \emph{Planck} consortia \citep{ERCSC11,ERCSC13,Wright09,WMAP9}. Sparse point-source removal is then applied on \emph{WMAP} data in Sect. \ref{sec:wmap9data}, code information is given in Sect. \ref{sec:repres}, and conclusions and perspectives are drawn in Sect. \ref{sec:ccl}.

\section{\label{sec:methods}Sparse point-source removal}
\subsection{\label{sec:modeling}Modelling of the sky}

We modelled  microwave-channel data over the full sky  $y \in \Rn{N_e}$ (i.e. the channel map is composed of $N_e$ real-valued pixels) composed of three components $\{x_1, x_2,x_3\}\in \Rn{N_e}$: the point sources, the extended compact sources, and the diffuse emissions. In this decomposition, the diffuse background is composed of the CMB itself and the synchrotron, free-free, dust diffuse emissions as well as all the sources with flux below the detection cut. The forward model can be cast as
\begin{equation}
\label{eq:model}
y = x_1 + x_2 + x_3+ n\;,
\end{equation}
where $n\in \Rn{N_e}$ is an additional noise, assumed to be a real-valued Gaussian random field (but not necessarily stationary). In the associated inverse problem, we need to estimate three unknowns from one equation, which is not possible without additional constraints/information. We now discuss these  assumptions and constraints on each of the components.

-- Point-source catalogues are available for \emph{WMAP} and \emph{Planck} data sets. Using these catalogues, $x_1$ can be modelled as a set of Dirac functions $\{\delta_p\}_p$ convolved by the instrumental beam $b$: $x_1 = \sum_{p=1}^{N_p} f_p \delta_p * b,$ where $N_p$ is the number of point sources that need to be removed and $f_p$ is the flux of the $p$th point source. The source fluxes are expected and enforced to be positive.

-- In case of extended compact sources, the  morphological information is weaker. Several extended sources were detected and flagged in the \emph{Planck} data set, mostly located close to the Galactic centre, while in the \emph{WMAP} data such sources were not considered (detection was performed outside a Galactic mask). In this work, we assumed that these emissions can be modelled through a sparse decomposition in the undecimated isotropic spherical wavelet dictionary \citep{starck:sta05_2}: $x_2 = \sum_{k_1}^{N_e} \sum_{j=1}^{N_s} w_{jk} \theta_{jk} $, where $N_s$ is the number of wavelet scales, $\theta_{jk}$ corresponds to a wavelet atom at scale $j$ and position $k$, and we have overall $N_w=N_e N_s$ wavelet coefficients at all scales. Assuming sparsity of this Galactic component, only a few wavelet coefficients $w_{jk}$ at selected locations $k$ close to the Galactic centre  are assumed significantly away from zero. 

-- The background emission $x_3$ is much more complex to model accurately. It is usually represented as a local low-order polynomial background (either baseline, or first order) because of its local smoothness, and in this work it is characterized by its sparsity in the spherical harmonic domain.  This assumption of sparsity in CMB applications is motivated in \citet{Starck2013}. $x_3$ is also assumed band-limited because of the beam effect.

These components can only be separated if the components are sparse enough in mutually incoherent dictionaries, such as a dirac basis and a spherical harmonic basis \citep{DonohoHuo,mca:Donoho-Elad}. Said differently, compact-source emissions are expected to decrease the sparsity level of the data measured on spherical harmonics, which will be the key ingredient driving the separation process in our approach. 

\subsection{\label{sec:inverse_problem}Associated inverse problem}

Using all these hypotheses and constraints, we consider the following inverse problem:
\begin{multline}
\label{eq:1}
\minimize_{ f\in \mathcal{C}, \; w\in\Rn{N_w}, \;a\in \mathcal{D}} \;\; \gamma|| a ||_1 + \beta||w||_1 \;\mathrm{s.t.}
\; || y -   (\mb{B}f + \mb{M}\mb{W} w + \mb{S} a )||_{2,\mb{\Sigma}} < \epsilon \;,
\end{multline}
where
\begin{itemize}
\item[-]  $f\in\Rn{N_p}$ is a vector containing the flux of the $N_p$ point sources,  $\mathcal{C}$ is the positive orthant of $\Rn{N_p}$ to enforce the positivity constraint, and $\mb{B} \in \Rn{N_e \times N_p}$ is a matrix operating on fluxes and implementing the local projection of the beam at the location of the point sources;
\item[-] $w= \left\{w_{jk}\right\}_{j,k}\in\Rn{N_w}$ is a vector containing the $N_w$ wavelet coefficients at all scales,  $\mb{W}$ represents the matrix implementing the spherical wavelet transform (containing each wavelet atom as a column), $\mb{M}$ corresponds to a binary mask equal to one where compact extended sources are expected. We promote sparsity of $x_2$ by making use of the $\ell_1$ norm of $w$ as a penalization term;
\item[-] $a \in \mathbb{C}^{N_l}$ contains the $N_l$  complex spherical harmonic coefficients of $x_3$ and $\mathcal{D}$ is the space of band-limited signals on the sphere (up to a chosen multipole $\ell_{max}$ depending on the resolution of the channel). $\mb{S} \in\mathbb{C}^{N_e\times N_l}$ is the matrix describing the orthogonal spherical harmonic transform:  $x_3 = \mathbf{S} a$.
Sparsity of  $x_3$ is enforced through the use of a  $\ell_1$ norm penalization term on $a$, which is computed as the sum of the modulus of the complex multipole as in the sparse inpainting described in \citet{Starck2013}
\item[-] $\mb{\Sigma}$ is the (non-stationary) noise covariance matrix and $||x||^2_{2,\mb{\Sigma}}=x^T \mb{\Sigma}^{-1} x$ denotes the square of the $\ell_2$ norm weighted by $\mb{\Sigma}^{-1}$.
\end{itemize}

The reconstructed point-source-free map $\widetilde{y}$ can be obtained by $\widetilde{y} = y -  \mb{B}f $.
The problem described in Equation \ref{eq:1} is a convex problem and is related to a constrained morphological component analysis \citep{starck:sta04,starck:sta05_4} or a basis pursuit denoising problem \citep{wave:chen98} with a deconvolution step. The principal advantage of this constrained formulation lies in having only a few hyperparameters to set, which are easily interpretable: the expected noise level $\epsilon$; the balance between $\ell_1$ norm in the wavelet and spherical harmonic domains. More explicitely, the higher the ratio  (${\gamma}/{\beta}$), the sparser the solution for $x_3$ is in the spherical harmonic domain, and conversely, the lower the ratio, the sparser the estimate of $x_2$ in the wavelet domain. Setting them equal would lead to penalize both non-sparse solutions for $x_2$ in the wavelet domain and $x_3$ in the spherical harmonic domain, as performed in morphological component analysis \citep{starck:sta04,starck:sta05_4}. Note that once the ratio is set, changing their modulus does not change the inverse problem.

The main difficulty here lies in controlling the interplay between sparsity and data fidelity constraints: how can we efficiently estimate a sparse solution (more precisely with minimal $\ell_1$ norm), knowing that many combinations of $x_1$, $x_2$ and $x_3$ can satisfy the data fidelity constraint?  

In the next section, we propose an algorithm to solve this problem.

\subsection{\label{sec:algorithm}Proposed algorithm}

Inverse problems as described by Equation~\ref{eq:1} can be solved using primal-dual approaches \citep[e.g.][]{Chambolle2011,Briceno11,Becker11,Combettes12}. The proposed algorithm was derived from \citet{Chambolle2011}; it requires only one application of the costly spherical harmonic and wavelet transforms and one application of their adjoint per iteration and does not require sub-iterations. The algorithm is set as follows:

\noindent\fbox{
\begin{minipage}[h]{0.9\linewidth}
\textbf{Morphological component analysis with a primal-dual approach (\bf{SPSR})}
\begin{itemize}
\item[1-] Choose $(a^0,f^0,w^0,t^0)\in \ \mathbb{C}^{N_l} \times \Rn{N_p}  \times \Rn{N_w}  \times \Rn{N_e}$ and  $\bar{a}^0=a^0$, $\bar{f}^0=f^0$, $\bar{w}^0=w^0 $. \\Choose also the hyperparameters $\gamma, \beta$ and $\tau, \sigma$ s.t. $\tau\sigma<\fracd{1}{3}$ (assuming normalized $||S||_2=||MW||_2=||B||_2 = 1$). 
\item[2-] Iterate ($n \geq 0$):
\begin{equation}
\hspace*{-0.5cm}
\left\{ 
\begin{array}{l}
r_d^{n}=t^n +\sigma\Sigma^{-1/2} (\mathbf{B}\bar{f}^n  + \mathbf{S}\bar{a}^n+ \\
\hspace{4cm} \mathbf{MW}\bar{w}^n - y)\\[0.3cm]
 t^{n+1}=
\left\{
\begin{array}{l}
0\; \mathrm{if}\; ||r_d^n||_2 \leq \sigma\epsilon\\[0.5cm]
(1-\fracd{\epsilon\sigma}{||r_d||_2}) r_d \;\;\mathrm{otherwise}
\end{array}
\right. \\[0.8cm]
 a^{n+1} = \mathcal{ST}_{\tau\gamma} \left( P_{\mathcal{D}} (a^n-\tau  \mathbf{S}^\dagger \mathbf{\Sigma}^{-1/2 \dagger} t^{n+1})\right)\\[0.3cm]
 f^{n+1}=P_{\mathcal{C}}\left (f^n - \tau  \mathbf{B}^\dagger  \mathbf{\Sigma}^{-1/2 \dagger} t^{n+1}\right)\\[0.3cm]
 w^{n+1}=\mathcal{ST}_{\tau\beta} (w^n -  \mathbf{W}^{\dagger}  \mathbf{M} \mathbf{\Sigma}^{-1/2 \dagger} t^{n+1})\\[0.3cm]
\bar{f}^{n+1}=2 f^{n+1} - f^n\\
\bar{a}^{n+1}=2 a^{n+1} - a^n\\
\bar{w}^{n+1}=2 w^{n+1}-w^n\\
\end{array}
 \right.
\end{equation}
\end{itemize}
\end{minipage}
}

where $P_{\mathcal{C}} $ is the projection onto the positive orthant (i.e. sets to $0$ negative fluxes) and $P_{\mathcal{D}}$ is the projection onto the set of considered band-limited signals (i.e. sets to $0$ all multipoles greater than the chosen $\ell_{max}$); $\mathcal{ST}_{\tau\beta}$ is the standard soft-thresholding operator applied component-wise:
\begin{equation}
\left[\mathcal{ST}_{\tau\beta} x\right]_i=x_i\;\left[1-\frac{\tau\beta}{|x_i|}\right]_+ \;,
\end{equation}
where $|x_i|$ is the complex modulus of $x_i$ for complex vectors. From \citet{Chambolle2011}, the sequence $((f^n,a^n,w^n),t^n)$ converges to a saddle point of the primal-dual problem with a restricted dual gap decreasing as $\mathcal{O}(1/n)$ (first-order method).  We initialised the algorithm with null images.

The sparse-source-removal (SPSR) algorithm has four parameters to set: $\beta$, $\gamma$, $\tau$, and $\sigma$. The last two are the primal and dual steps, respectively, and drive the convergence speed of the algorithm. The ratio of the first two parameters is the hyperparameter that controls the balance in the sparsity of the decomposition of the extended compact sources and the diffuse background as discussed in the previous section. The choice of their modulus also affects the convergence speed of the algorithm: a high modulus leads to a slowly built sparse approximation, or said differently, the algorithm would provide very sparse solutions at the beginning of the algorithm that are far from satisfying the data fidelity constraint,
however;  a low modulus, in contrast, leads to building approximations that satisfy the data fidelity constraint early in the iterations, but are not as sparse. In practice, several values need to be experimentally tested to obtain an algorithm with reasonable convergence speed and to derive the required number of iterations.

To assess the relative performance of the proposed approach, flux estimates obtained with SPSR were compared with the low-order polynomial fitting approach \citep{WMAP9,ERCSC11,ERCSC13}. Fluxes were estimated in a $3.5\sigma$ region as previously recommended in \citet{Wright09} using a Levenberg-Marquardt algorithm with local $\chi^2$ minimization (using the C++ library ALGLIB (www.alglib.net), Sergey Bochkanov and Vladimir Bystritsky). The local background was either modelled with a baseline or a first-order polynomial (\tb{FIT-C} and \tb{FIT-L} in the following) and a positivity constraint was applied on the fluxes. Clustered point sources with overlapping fitting regions were jointly fitted for. However, in practice, such situations were scarce in the \emph{WMAP} simulations reported.

\section{Results on synthetic \emph{WMAP} simulations}
\label{sec:results}

\subsection{Planck Sky Model simulations at WMAP frequencies}

The \emph{Planck Sky Model} (PSM) software \citep{Delabrouille13} was employed to simulate \emph{WMAP}-like data. Each one of the five \emph{WMAP} frequency channels included the following:
\begin{itemize}
\item[--] A diffuse component, comprised of a Gaussian CMB, generated from a six-parameter $\Lambda$-CDM model (with default values from WMAP 7-year data combined with BAO and the Hubble constant measurements, and with added  $C_\ell$ lensing) and synchrotron, free-free, thermal-dust and spinning-dust emissions, simulated with default PSM parameters.  
\item[--] A compact component,  constituted by radio, infrared, and strong ultra-compact HII region emissions, as well as a far-infrared background (model jgn2005 in the PSM). 
More specifically, radio sources are essentially derived from observations conducted between $0.85$ GHz and $4.85$ GHz. Each one of these $\sim2$ million simulated sources possessed its own spectrum, composed of power laws with spectral indices varying across pre-defined bands of electromagnetic frequencies. As reported in \citet{Delabrouille13},  this catalogue is expected to faithfully reproduce the clustering properties of the radio sources and the observed/modelled source number counts at 5 and 20 GHz.
In the PSM, the infrared sources were derived from the IRAS point-source catalogue and faint-source catalogue, with assumed modified blackbody emissions to extrapolate the flux at lower frequencies. Simulated sources were also added near the Galactic plane and in IRAS gaps to obtain the same mean surface density as a function of flux (down to $80\textrm{mJy}$) as in the regions well covered by IRAS. 
Emissions from ultra-compact  HII regions were also derived from IRAS, with flux extrapolated according to a modified blackbody fit with added low-frequency  flux with a free-free spectral index when a radio-counterpart is found. Note that compact emissions from the Galaxy were also simulated as part of diffuse emissions.
Finally, the simulated far-infrared background is composed of realistic distribution of clustered point sources with assumed flux-dependent spectral indices and reproduced \emph{Planck} observations reasonably well. 
\item[--] Noise modelled as a non-stationary Gaussian random field, with variance in each pixel derived from the \emph{WMAP} 9-year hit maps; the channel bandpasses were modelled as diracs located at the centre frequency, and the beams were assumed Gaussian with a full-width at half maximum given by the beam size provided by the \emph{WMAP} consortium. 
\end{itemize} 

\begin{figure*}[htb]
   \begin{center}
   \begin{tabular}{c}
   \includegraphics[height=5cm]{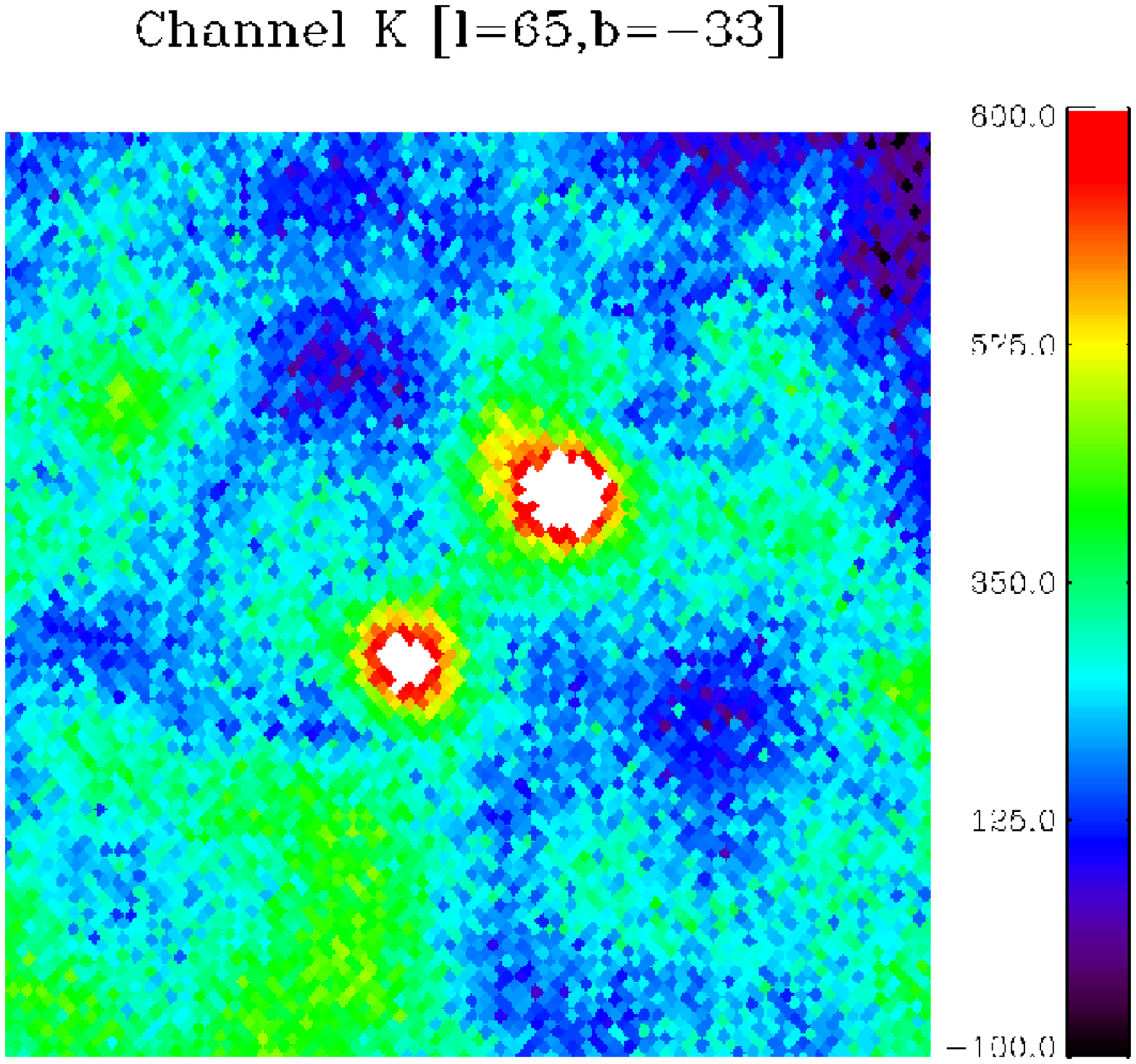}
   \includegraphics[height=5cm]{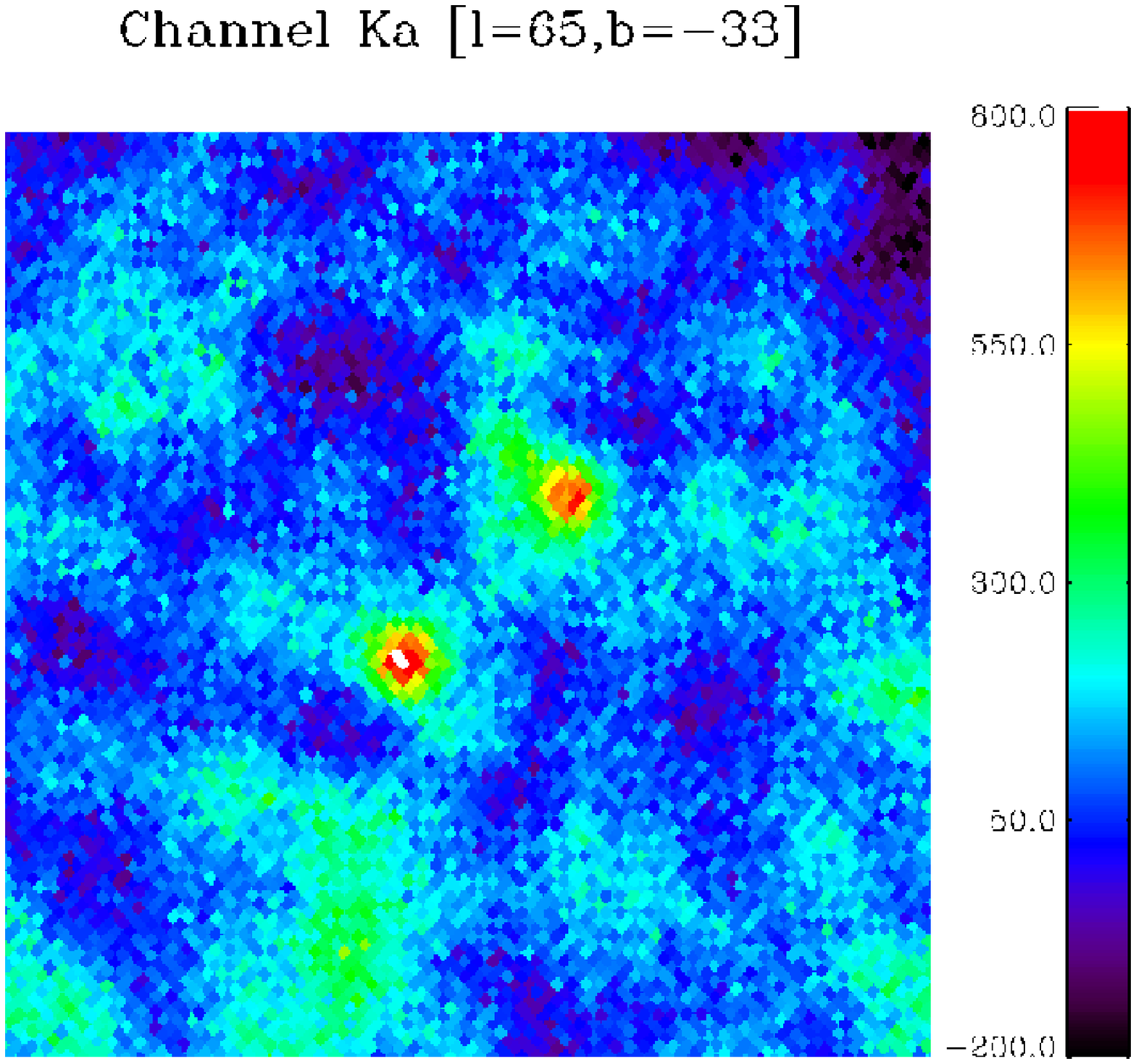}
   \includegraphics[height=5cm]{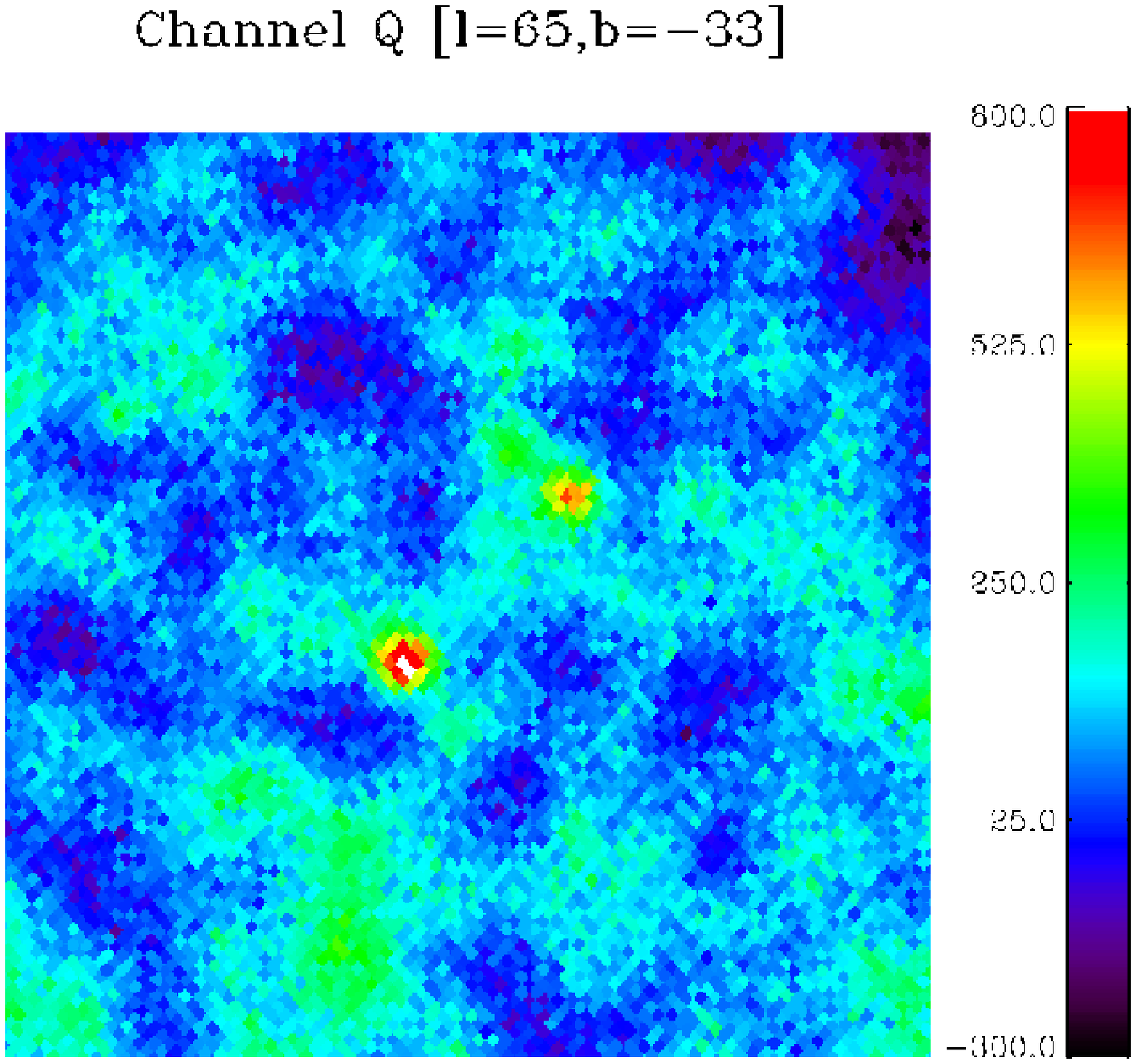}\\
    \includegraphics[height=5cm]{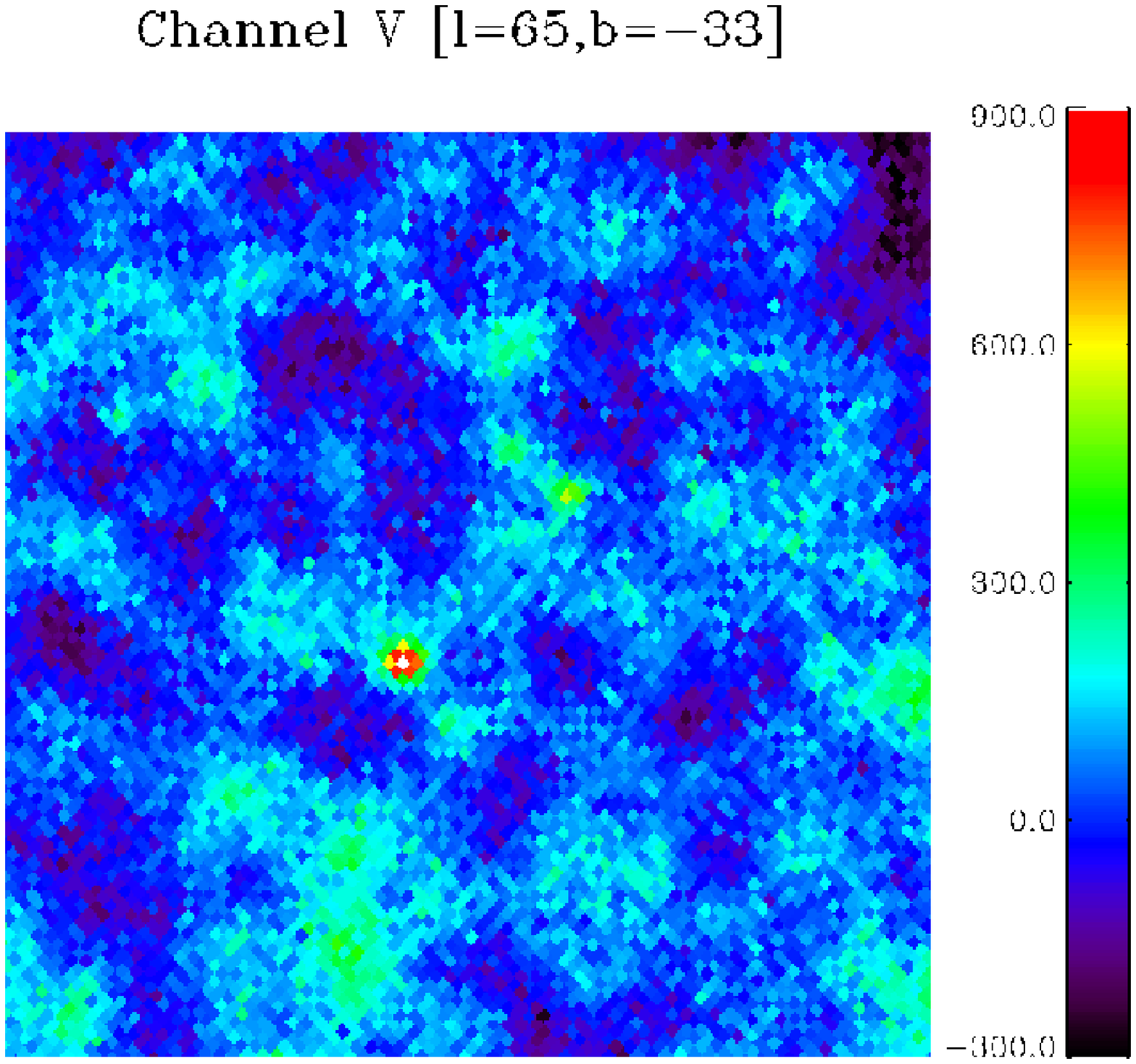}
   \includegraphics[height=5cm]{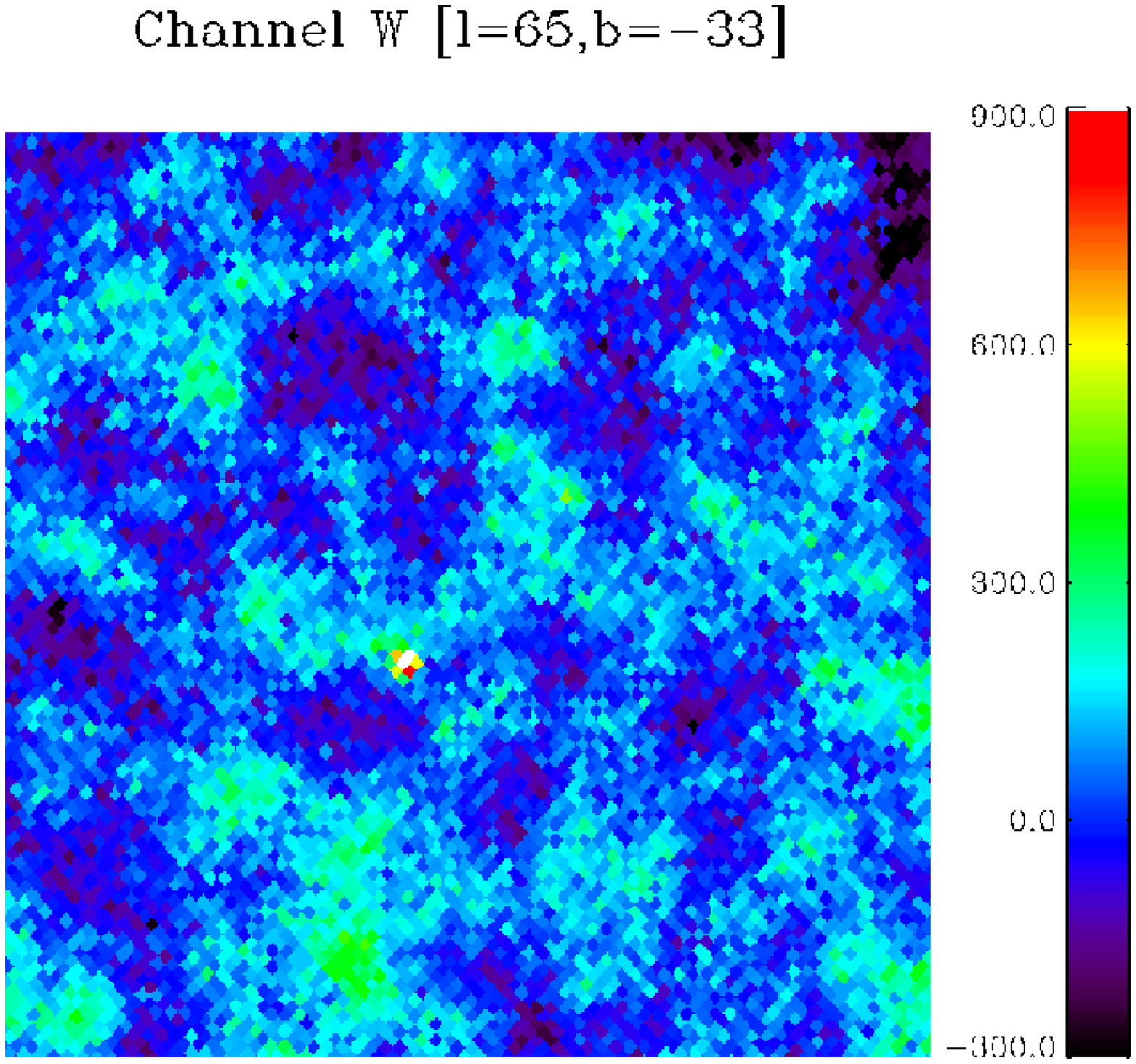}\\
   \end{tabular}
   \end{center}
   \caption[example] 
   { \label{fig:patchsimu} Patch of the simulated sky centred on two detected point sources at the \emph{WMAP} wavelengths.}
    \end{figure*}

Overall, the PSM simulations reflect the complexity in estimating the strong point-source fluxes (which is essential to test the robustness of the proposed approach), with complex diffuse background, unresolved sources with clustering properties below the detection limit, strong compact emissions in the Galactic plane, and non-stationary noise. A more comprehensive description of the PSM is provided in \citet{Delabrouille13}.
A patch extracted from these \emph{WMAP} 9-year simulated data is represented in Figure \ref{fig:patchsimu} at each of the five \emph{WMAP} frequency channels to illustrate the complexity of the source separation, because the resolution, noise, and background change in each of the channels. At low-frequency channels, the background fluctuates at the scale of the point sources, and noise contribution to the flux estimates are lower than in the high-frequency channels, where noise becomes important at the scale of the source, as reported in \citet{Wright09}.

\subsection{Flux recovery in noise-limited catalogue}

\subsubsection{Catalogue generation}
First, a catalogue of point sources was independently generated from the simulations for each \emph{WMAP} channel by retaining the sources with flux above $5\sigma$ of the estimated radiometer noise \citep{Wright09}. The number of sources retained for each channel and the flux cut are displayed in Table~\ref{tab:pointcatfluxcut}. This scenario reflects an optimistic situation where only noise would be a limiting factor for source detection (and not the background). Consequently, it allows us to investigate how well point sources can be estimated even when their flux is similar to the background levels. Note that the cut in flux increases with frequency (because of the increasing level of the noise) and our sources are mainly radio sources (i.e. with flux decreasing with increasing frequency). Consequently, fewer sources are detected at higher frequency channels.

\begin{table}[h]
\caption{\bf Characteristics of the point-source catalogue generated from simulations.} 
\label{tab:pointcatfluxcut}
\begin{center}       
\begin{tabular}{|l|l|l|} 
\hline
\rule[-1ex]{0pt}{3.5ex}  Channel & Flux Cut (mJy) & Source Number  \\
\hline
\rule[-1ex]{0pt}{3.5ex}  K & 111 & 1094  \\
\rule[-1ex]{0pt}{3.5ex}  Ka & 183 &  1093 \\
\rule[-1ex]{0pt}{3.5ex}  Q & 217 & 1037  \\
\rule[-1ex]{0pt}{3.5ex}  V & 383 &  683  \\
\rule[-1ex]{0pt}{3.5ex}  W & 687 &  330 \\
\hline 
\end{tabular}
\end{center}
\end{table} 

\subsubsection{SPSR working conditions}

The working conditions of SPSR were chosen as follows: the noise hyperparameter $\epsilon$ corresponds in the following to the 95th percentile of a $\chi^2(N_e)$ distribution according to the whitened noise statistics. For all channels the hyperparameter ratio  was set to $\beta/\gamma=1$  to favour both sparsity of $x_2$ in the wavelet domain and $x_3$ in the spherical harmonic domain, as discussed before.

\begin{figure*}[htb]
  \begin{center}
  \begin{tabular}{c}
   \includegraphics[height=7cm]{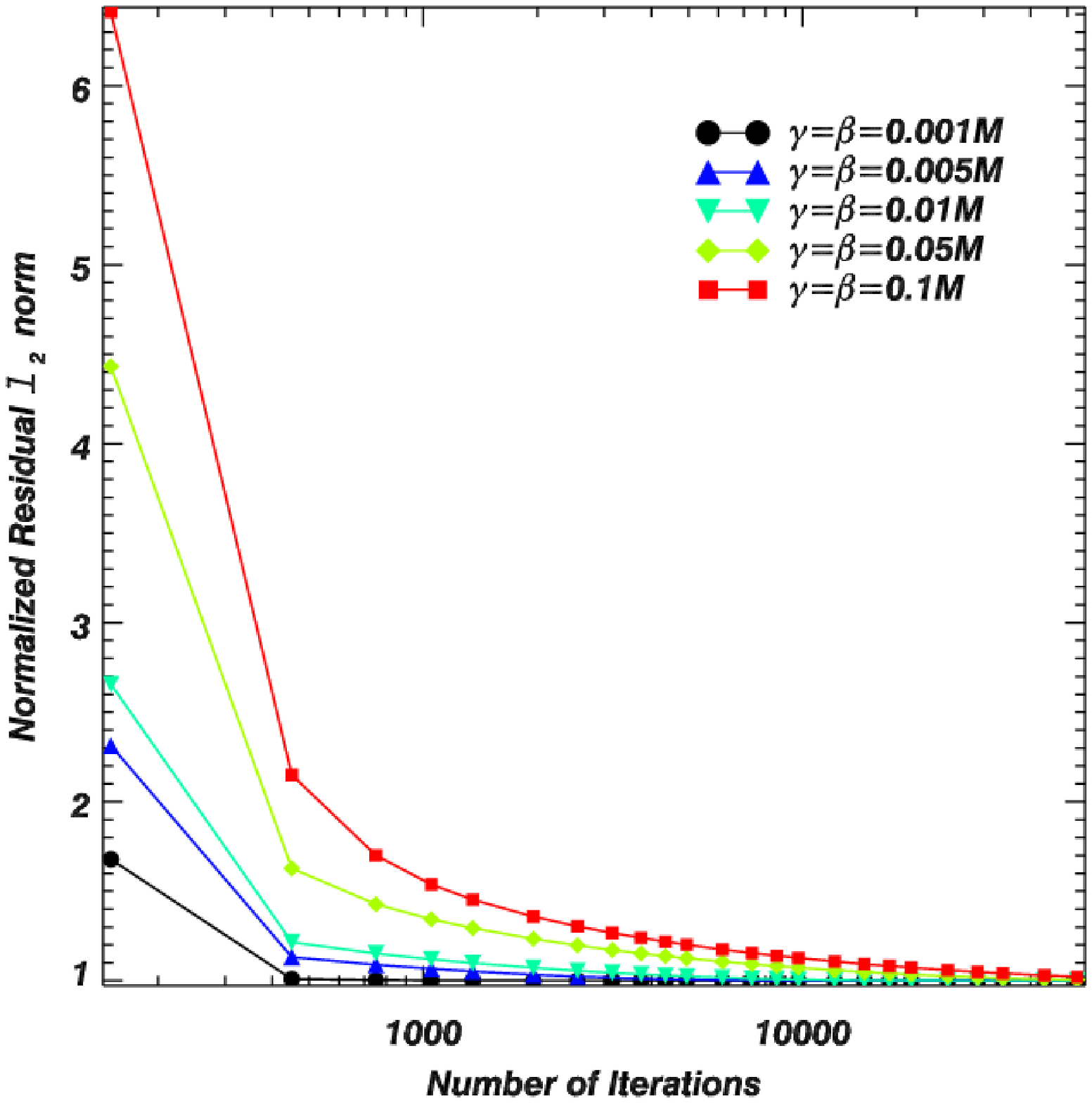}
   \hspace{1cm}
   \includegraphics[height=7cm]{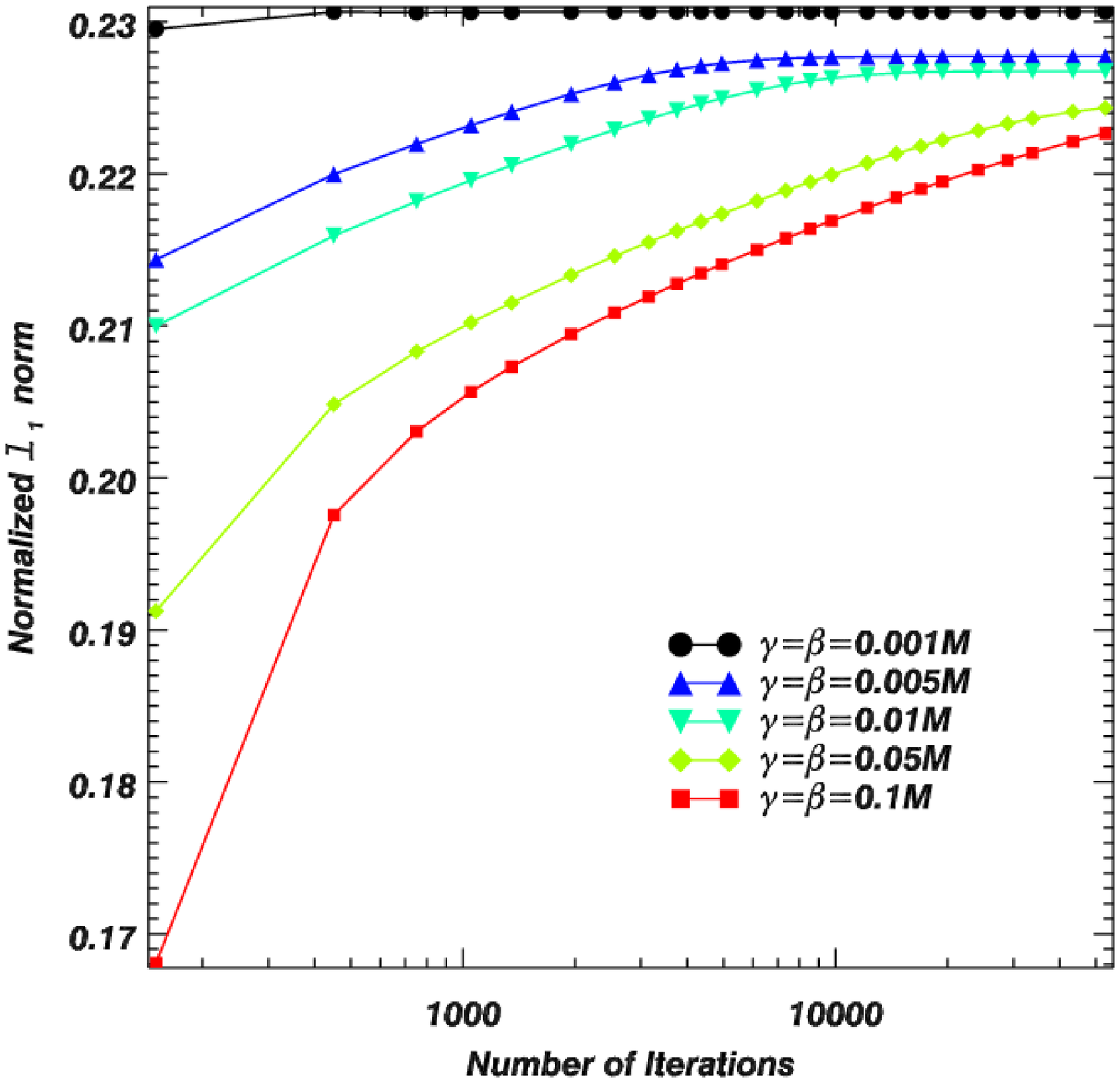}
   \end{tabular}
   \end{center}
   \caption[example] 
   { \label{fig:parameters_with_iterations} 
Evolution with the number of iterations of the normalized $\ell_2$ norm of the residuals (left) and normalized $\ell_1$ norm of the diffuse and compact component for channel Ka and for various values for the modulus of $\beta=\gamma$. Reaching the upper limit of the data-fidelity constraint leads to a normalized residual $\ell_2$ norm of $1$. The $\ell_1$ norm was normalized to the $\ell_1$ norm of the input data in direct space (sum of the absolute values of the pixels in the frequency map). $M$ corresponds to the highest modulus of the spherical harmonic coefficients in the input image.}
   \end{figure*}

A range of values for the modulus  of $\beta=\gamma$  was first tested and turned out to be crucial for the convergence speed of the algorithm. This is illustrated in Figure~\ref{fig:parameters_with_iterations}, where SPSR was run with a high number of iterations for channel Ka. Setting the modulus too high (or too low)  leads to many iterations necessary to fulfill the data-fidelity constraint (or minimizing the $\ell_1$ penalty for components that satisfy the data-fidelity term). For instance, for values of  $\beta=0.001 M$ (or $\beta=0.005M$) the data-fidelity constraint is satisfied in the first one thousand iterations, but the sparsity-penalty term decreases very slowly and has not converged after 50000 iterations. Conversely, $\beta=0.1 M$  (or $\beta=0.05M$)  leads to very sparse approximations that do not satisfy the data-fidelity constraint even after 50000 iterations. We chose the value $\beta=0.01 M$ for all channels, which led to the best convergence properties among the tested values.

\begin{figure*}[htb]
   \begin{center}
   \begin{tabular}{c}
   \includegraphics[height=7cm]{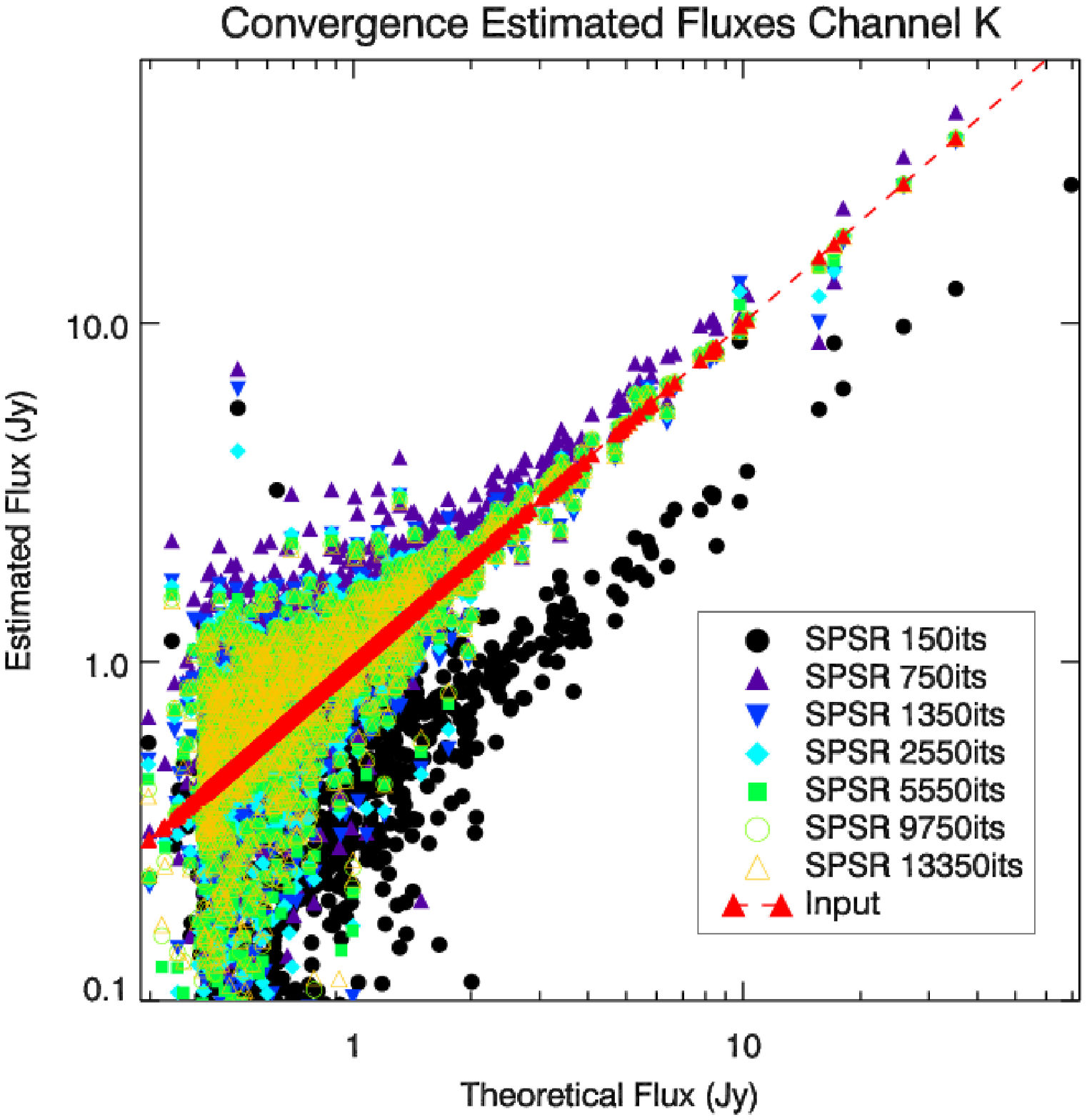}
   \hspace{1cm}
   \includegraphics[height=7cm]{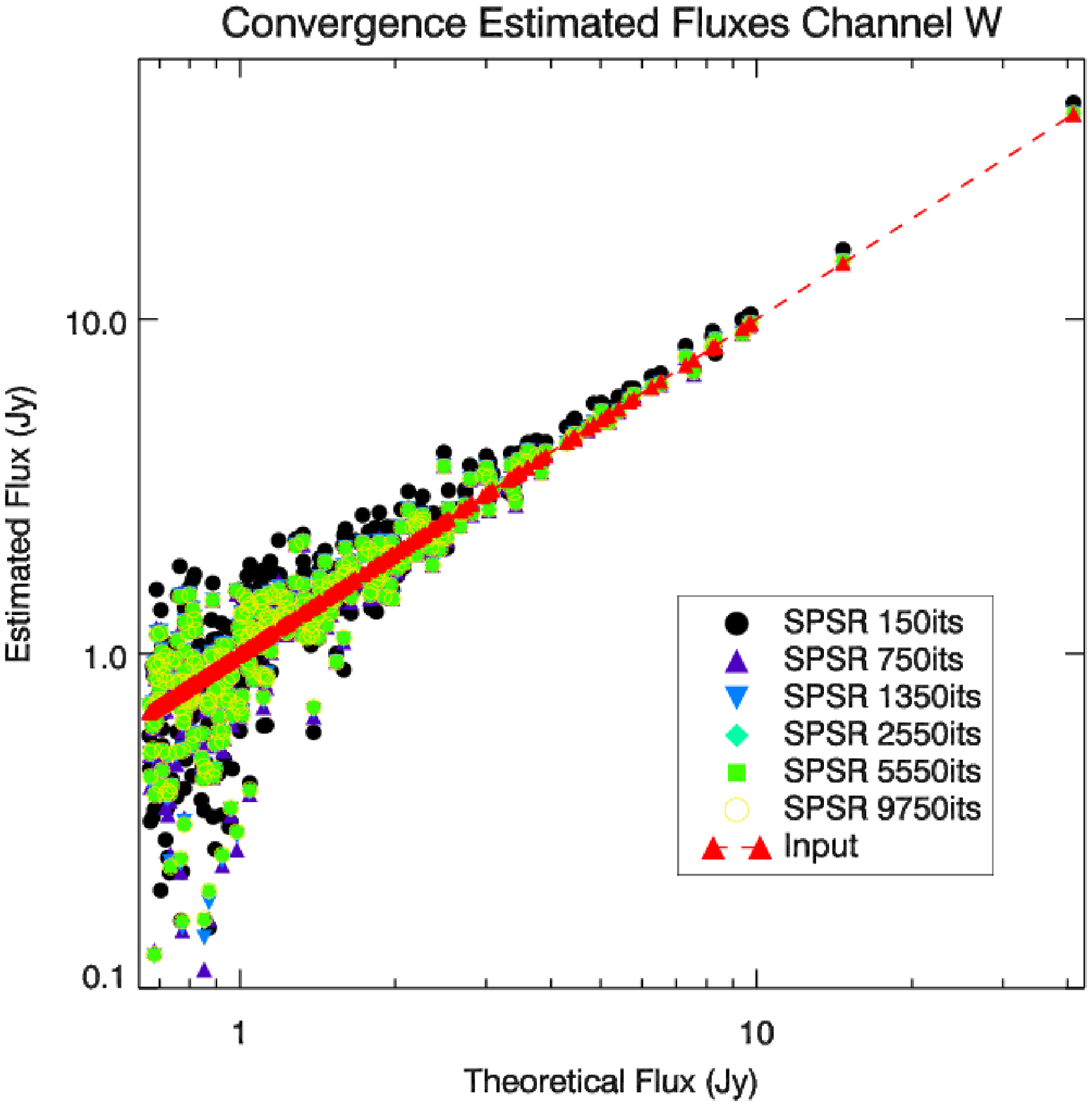}
   \end{tabular}
   \end{center}
   \caption[example] 
   { \label{fig:convspeed} 
Flux recovery with the proposed approach as a function of the number of iterations. After 13350 (1350) iterations, the flux estimates have converged for channel K (W). Note that the fluxes are displayed in logarithmic scales to visualize the convergence speed for both low- and high-flux sources.}
   \end{figure*}

Finally, the number of iterations needs to be set. In addition to the modulus of $\beta=\gamma$, the convergence speed of the algorithm depends on the conditioning of the beam matrix $\mb{B}$. The number of iterations therefore varies from channel to channel, requiring more iterations for lower resolution. The flux estimates at several stages of the proposed algorithm are illustrated in Figure~\ref{fig:convspeed} for channels K and W, which correspond to the lowest and highest resolution and to the highest and lowest noise, respectively.  In practice, 13350 iterations were chosen for channel K and 9750 iterations for all other channels as a compromise between processing time and level of convergence. We also ensured that the statistics measured were marginally changing with respect to the relative difference between methods when reaching this number of iterations. Generally, the large number of iterations in both cases is caused by the absence of regularity in the inverse problem of Equation \ref{eq:1}, which leads to first-order algorithms such as SPSR. To give a pratical example of the computational cost of SPSR, 9750 iterations for the channel Ka represent $\sim 80$ hours of processing in a cluster with 20 Intel Xeon 2 GHz processors, essentially spent in the spherical harmonic and wavelet transforms.

\subsubsection{Flux recovery and point-source subtraction}

We compared point-source flux recovery in different flux bands for the FIT-C, FIT-L, and SPSR approaches in the simulated channels. The bias, root-mean square errors (RMSE), and mean absolute deviation (MAD) (less sensitive to outliers) are displayed in Table~\ref{tab:RMSE_fluxcut}. For all methods, the RMSE or MAD varies from channels K to W as a consequence of increasing noise contribution and decreasing background fluctuations at the scale of the point sources, as previously observed in \citet{Wright09}. When the noise contribution to flux uncertainty is predominant (in channel W), FIT-C and FIT-L have generally lower biases, RMSE, and MAD than SPSR. Note, however, that the improvement is very small (at most $6\%$ difference in RMSE, bottom row of Table~\ref{tab:RMSE_fluxcut}). Therefore, it seems that in the noise-dominating case, fitting a few local parameters for the diffuse background, as in FIT-C and FIT-L, gives slightly more robust flux estimates than SPSR, where many more global parameters are estimated. 
The situation is the opposite when background fluctuations are the foremost contribution as in the first four channels: SPSR now leads to much lower RMSE and MAD than FIT-C or FIT-L, as well as lower biases (for the first three channels). In particular,  point sources with flux lower than $1\textrm{Jy}$ are better recovered by about $12\% $, $26\% $, $32\% $ and $8\% $ in terms of RMSE for channels K to V, with much lower biases for the first three channels (more than $70\textrm{mJy}$ better for channel $K,$ for instance).

\begin{table*}[htb]
\caption{Statistics on the recovered flux for each channel for the three approaches and for various flux bands (in mJy). The best results are displayed in bold.} 
\label{tab:RMSE_fluxcut}
\begin{center}       
\begin{tabular}{|l|l|l|l|l|l|l|l|l|l|l|} 
\hline
\rule[-1ex]{0pt}{3.5ex} & & \multicolumn{3}{c|}{Flux $<$ 1 Jy} & \multicolumn{3}{c|}{1 Jy $\leq$ Flux $<$ 5 Jy} & \multicolumn{3}{c|}{Flux $\geq$ 5 Jy} \\
\rule[-1ex]{0pt}{3.5ex} Channel &Method  & Bias & RMSE & MAD & Bias & RMSE & MAD & Bias & RMSE &MAD \\
\hline
\rule[-1ex]{0pt}{3.5ex}  \multirow{4}{*}{K} & FIT-C & 114.1 & 392.9 & 292.6 & 112.4 & 384.9 & 292.2 & 67.6  & 650.71 &   404.92\\
\rule[-1ex]{0pt}{3.5ex} & FIT-L & 114.8 & 380.9 & 286.0 & 96.0  & 471.3 & 298.7 & 90.0 & 518.5 &   317.9\\
\rule[-1ex]{0pt}{3.5ex}  & SPSR & \tb{41.5} & \tb{334.8}& \tb{260.4} & \tb{47.1} & \tb{362.6} & \tb{272.6} & \tb{-16.9} & \tb{364.7}  &  \tb{247.0} \\
\hline
\rule[-1ex]{0pt}{3.5ex}  \multirow{4}{*}{Ka} & FIT-C & 70.8 & 357.4 & 283.0 & 50.4 & 392.6 & 312.7 & -26.9  & 548.1 &   356.3\\
\rule[-1ex]{0pt}{3.5ex} & FIT-L & 69.4 & 349.4 & 281.1 & 38.7  & 468.4 & 313.7 & 30.6 & 508.5 &   378.9\\
\rule[-1ex]{0pt}{3.5ex}  & SPSR & \tb{25.7} & \tb{259.0} & \tb{205.8} & \tb{6.6} & \tb{292.5} & \tb{220.9} & \tb{1.3} & \tb{369.5}  &  \tb{ 289.0} \\
\hline
\rule[-1ex]{0pt}{3.5ex}  \multirow{4}{*}{Q} & FIT-C & 36.8 & 319.9 & 259.0 & 29.5 & 356.0 & 284.8 & -38.1  & 380.5 &   292.2\\
\rule[-1ex]{0pt}{3.5ex} & FIT-L & 38.5 & 332.2 & 258.9 & 10.8  & 400.8 & 290.0 & \tb{13.0} & 310.10 &   249.2\\
\rule[-1ex]{0pt}{3.5ex}  & SPSR & \tb{11.0} & \tb{217.3} & \tb{171.4} & \tb{10.2} & \tb{254.6} & \tb{184.7} & 21.9 & \tb{242.8}  &   \tb{192.8} \\
\hline
\rule[-1ex]{0pt}{3.5ex}  \multirow{4}{*}{V} & FIT-C & -9.1 & 287.7 & 235.7 & 34.7 & 296.5 & 241.5 & 37.0  & 281.5 &   200.1\\
\rule[-1ex]{0pt}{3.5ex} & FIT-L & \tb{-0.55} & 321.3 & 240.4 & 16.0  & 353.8 & 245.0 & 42.6 & 304.5 &   214.4\\
\rule[-1ex]{0pt}{3.5ex}  & SPSR & -1.41 & \tb{263.9} & \tb{208.9} &\tb{13.8} & \tb{265.2} & \tb{208.2} & \tb{1.8} & \tb{231.3}  &  \tb{167.5} \\
\hline
\rule[-1ex]{0pt}{3.5ex}  \multirow{4}{*}{W} & FIT-C & -56.1 & 281.7 & 224.2 & \tb{18.0} & 266.8 & 207.6 & \tb{2.6}  & 346.8 &   254.3\\
\rule[-1ex]{0pt}{3.5ex} & FIT-L & \tb{-54.7} & \tb{280.1} & \tb{222.8} & 20.0  & \tb{265.9} & \tb{206.9} & 13.6 & 283.6 &   224.4\\
\rule[-1ex]{0pt}{3.5ex}  & SPSR & -62.0 & 296.5& 229.6 & 30.4 & 273.8 & 208.6 & 11.7 & \tb{257.6}  &  \tb{ 202.4} \\
\hline
\end{tabular}
\end{center}
\end{table*} 

\begin{figure*}[htb]
   \begin{center}
   \begin{tabular}{ccc}
   \includegraphics[width=4.8cm]{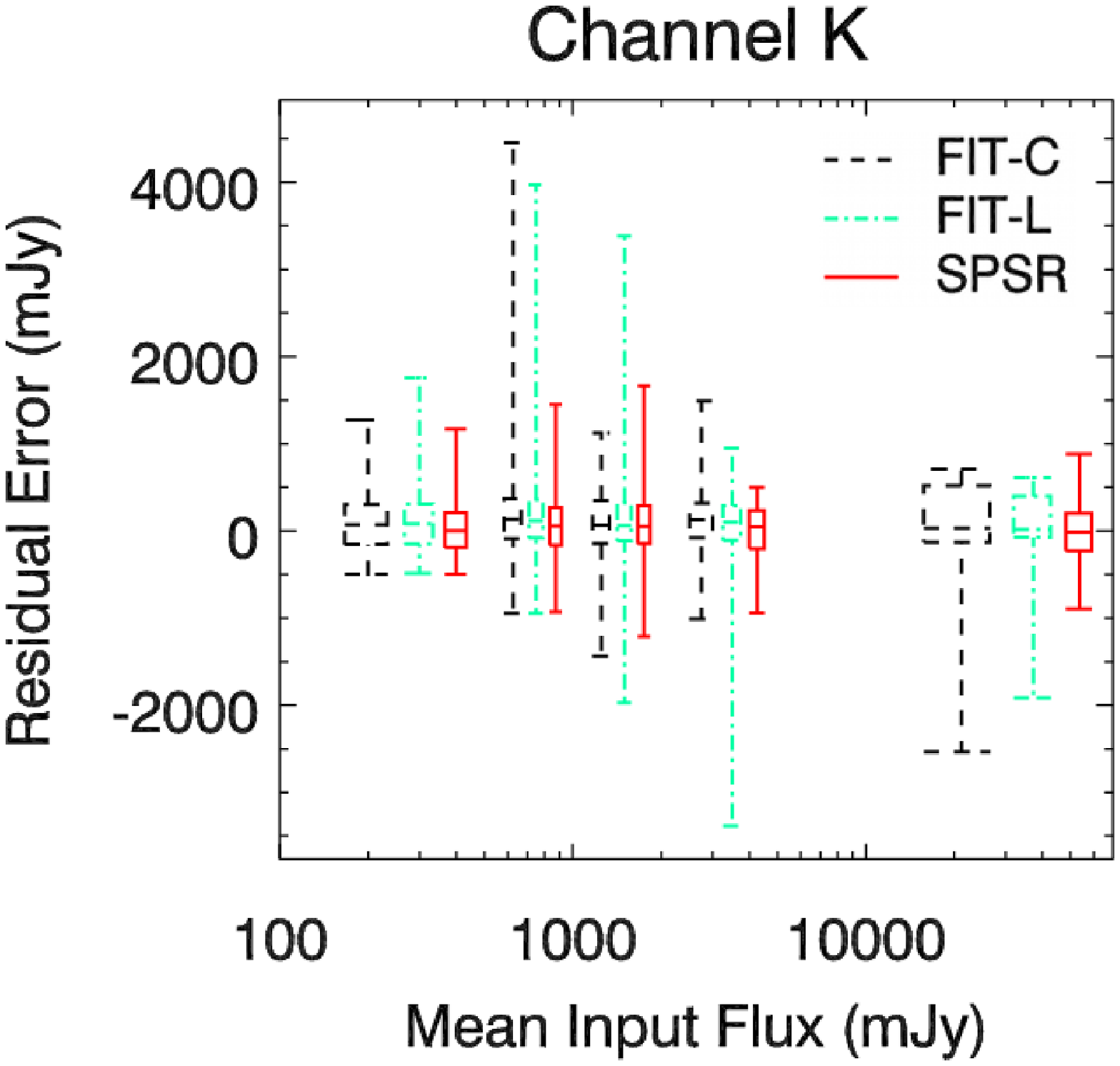} &   \includegraphics[width=4.8cm]{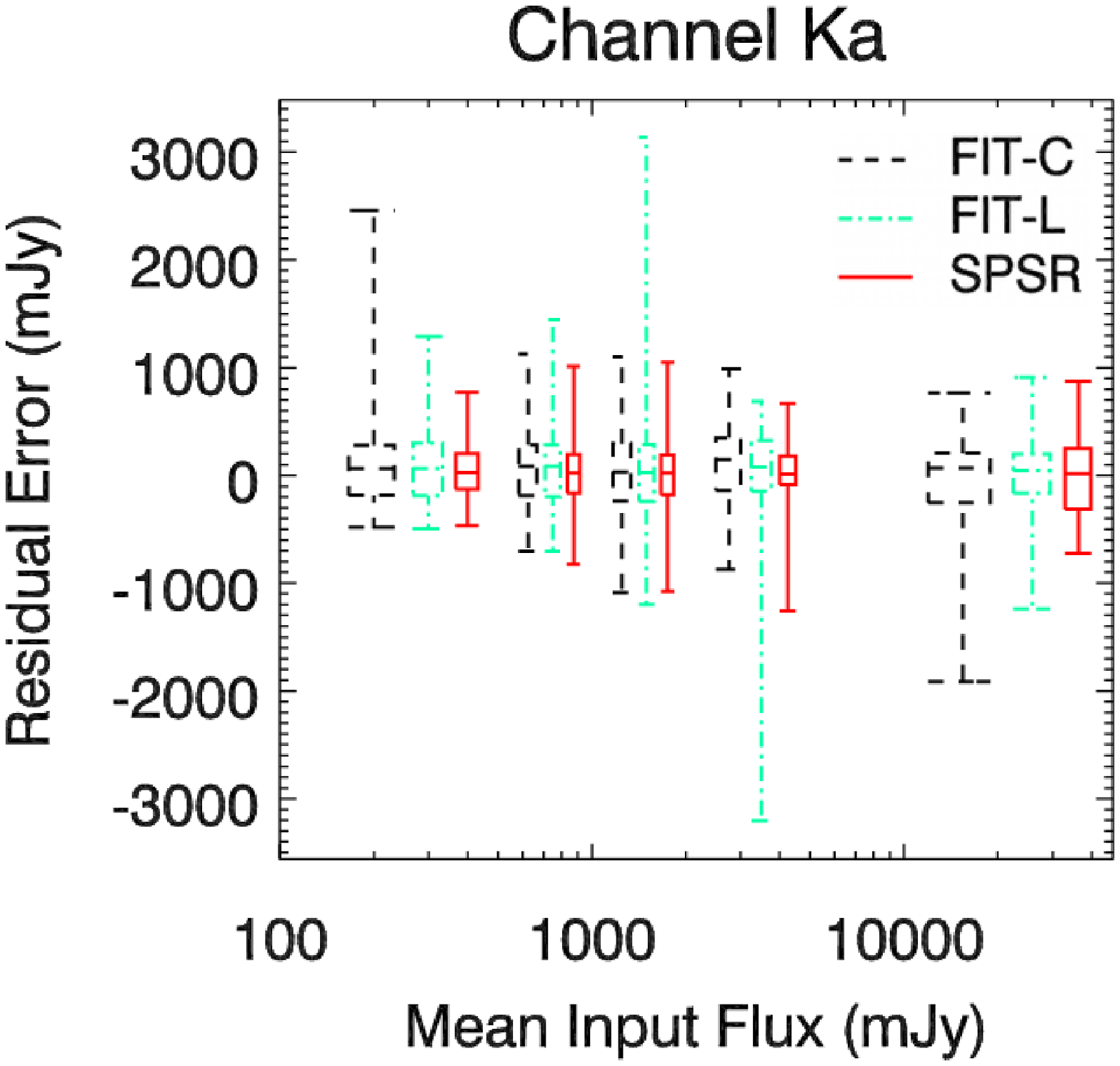} & \includegraphics[width=4.8cm]{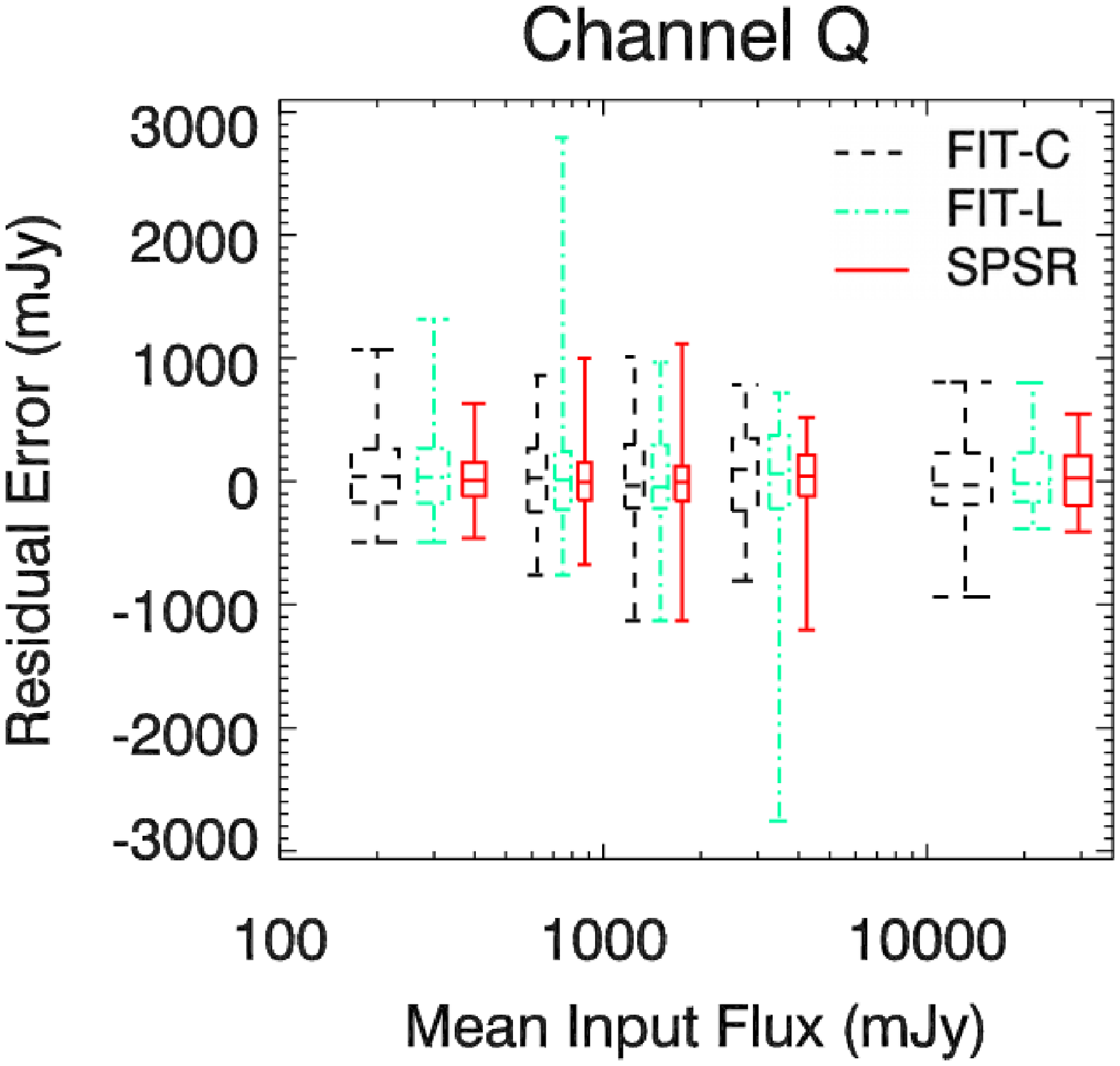}
   \end{tabular}
   \begin{tabular}{cc}
      \includegraphics[width=4.8cm]{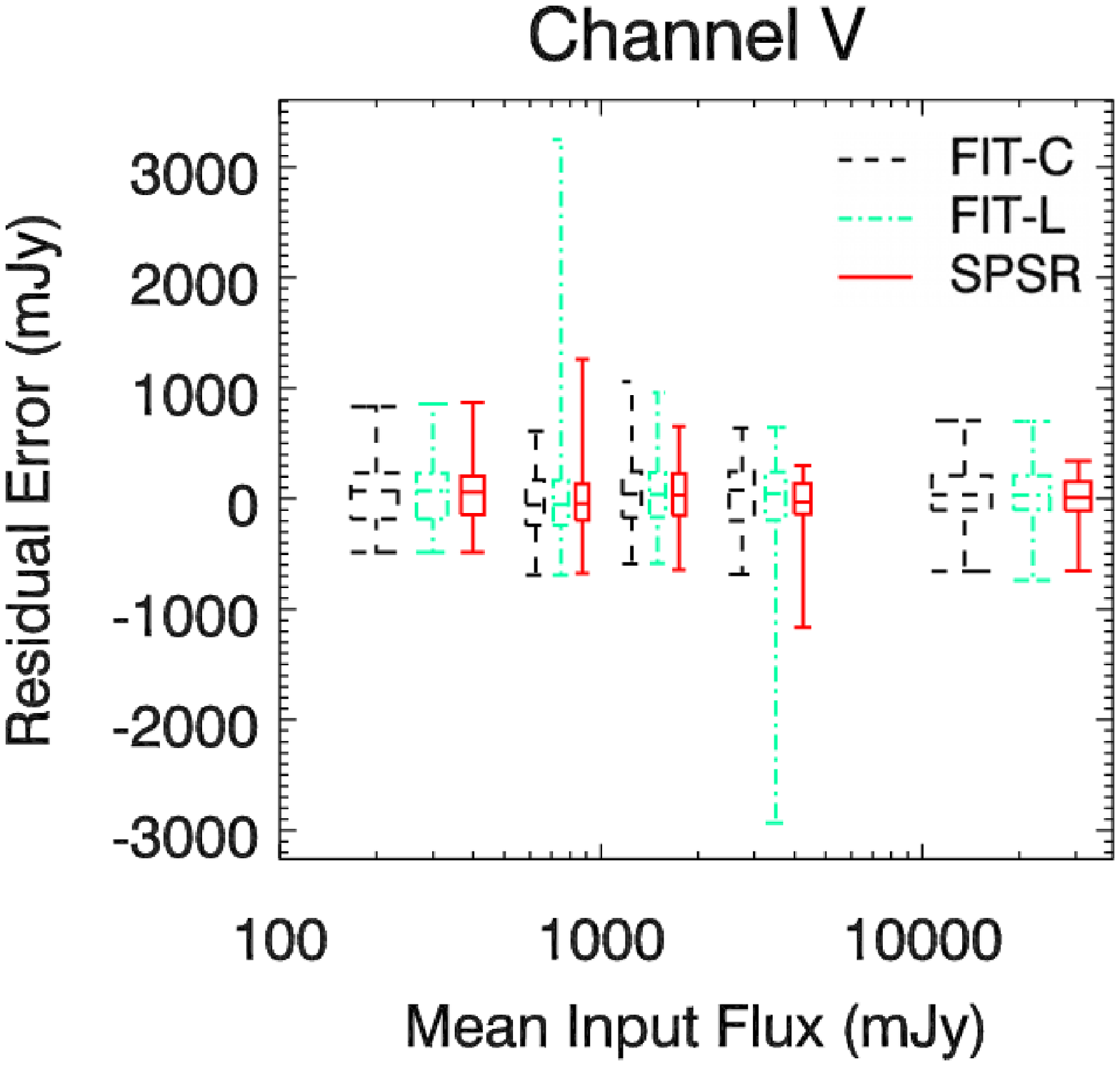} &   \includegraphics[width=4.8cm]{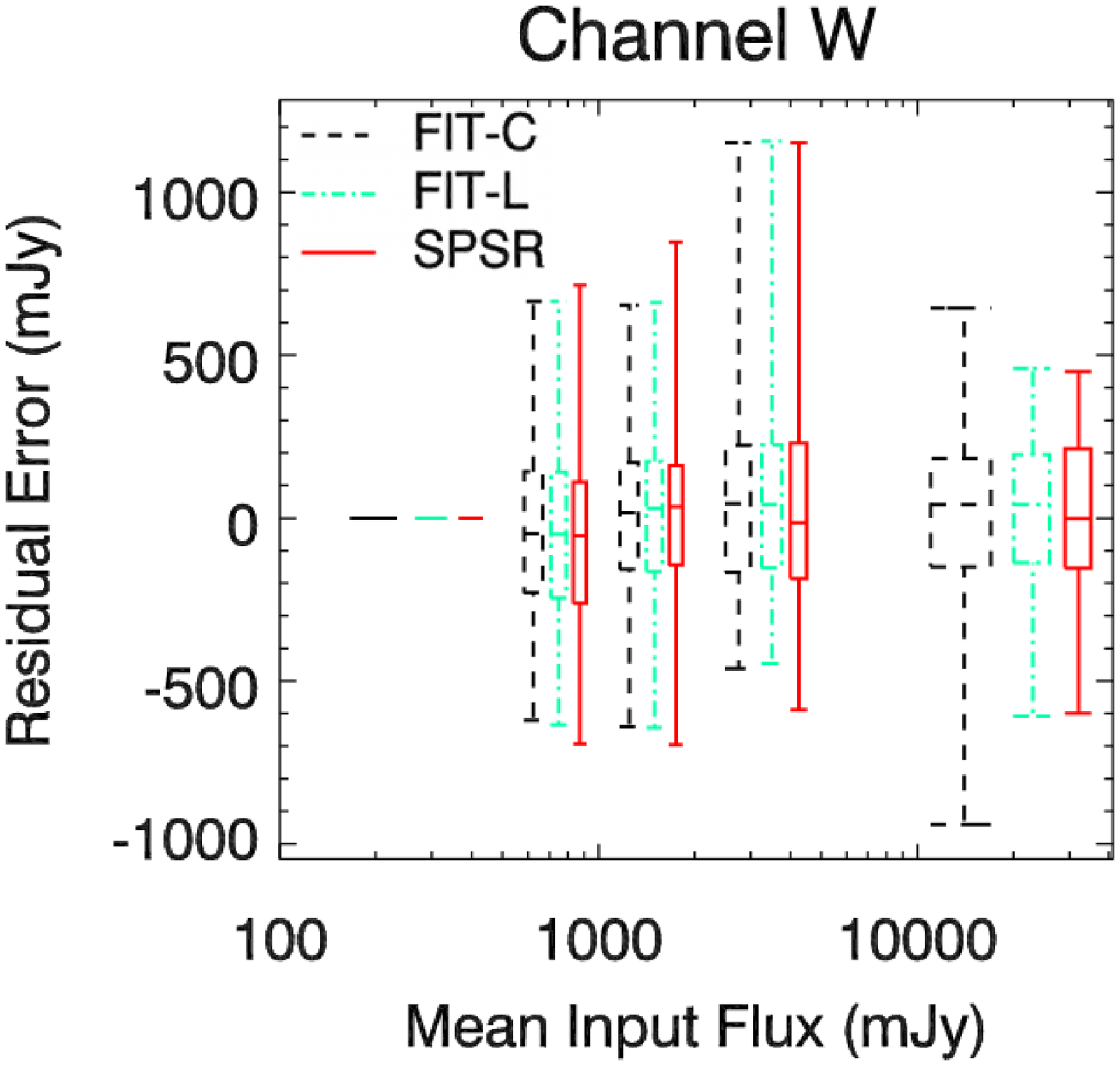}\\
   \end{tabular}
   \end{center}
   \caption[example] 
   { \label{fig:Boxplots} Statistics on error flux using the various approaches for the five channels for five  different flux bands (flux $< 500$ mJy, $500$ mJy $\leq $ flux $< 1$ Jy, $1$ mJy $\leq $ flux $< 2$ Jy, $2$ Jy $\leq $ flux $< 5$ Jy, flux $\geq 5$ Jy).  Quartiles and extreme values are plotted. }
   \end{figure*}

\begin{figure*}[htb]
   \begin{center}
   \begin{tabular}{ccc}
   \includegraphics[width=4.8cm]{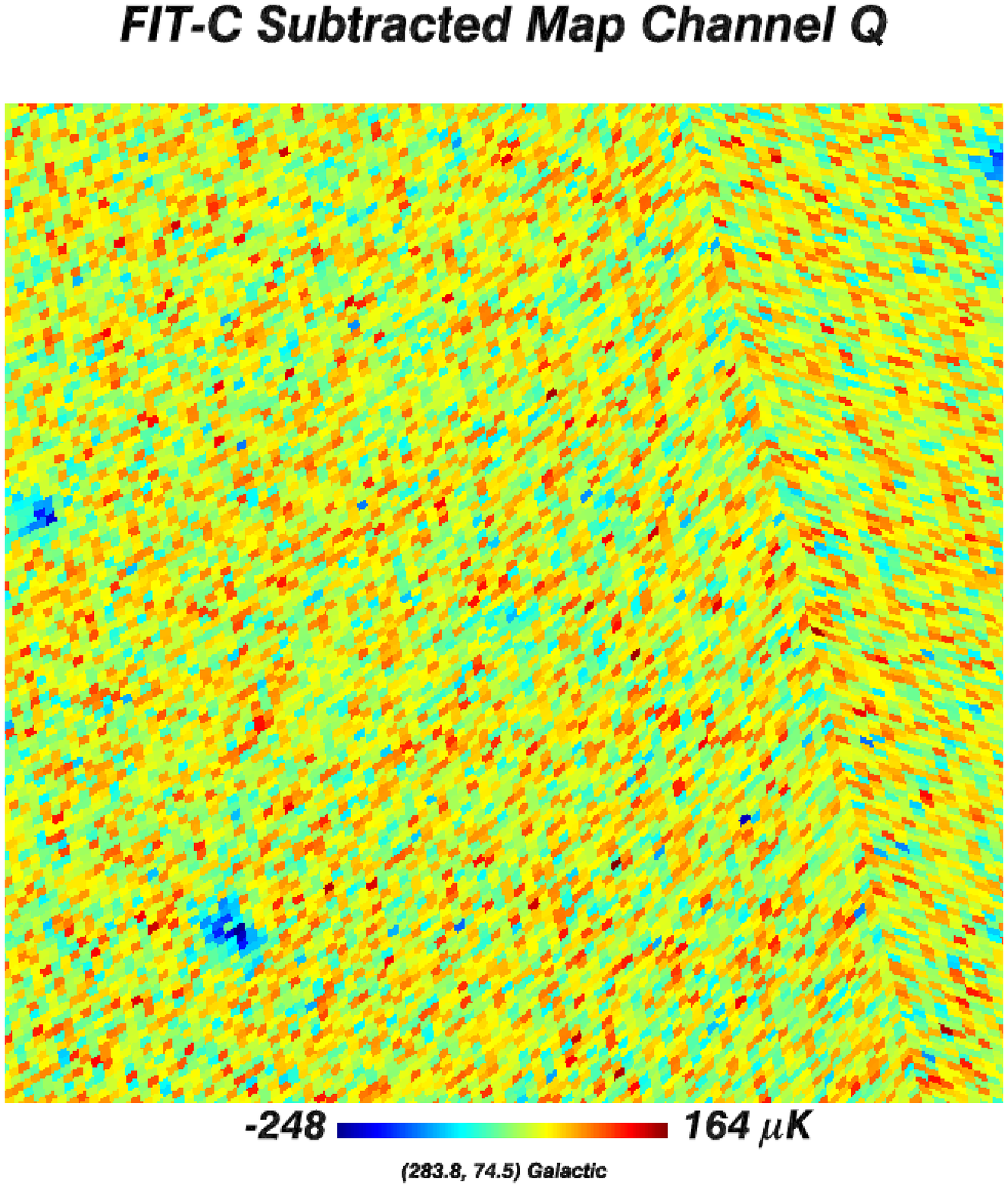} &   \includegraphics[width=4.8cm]{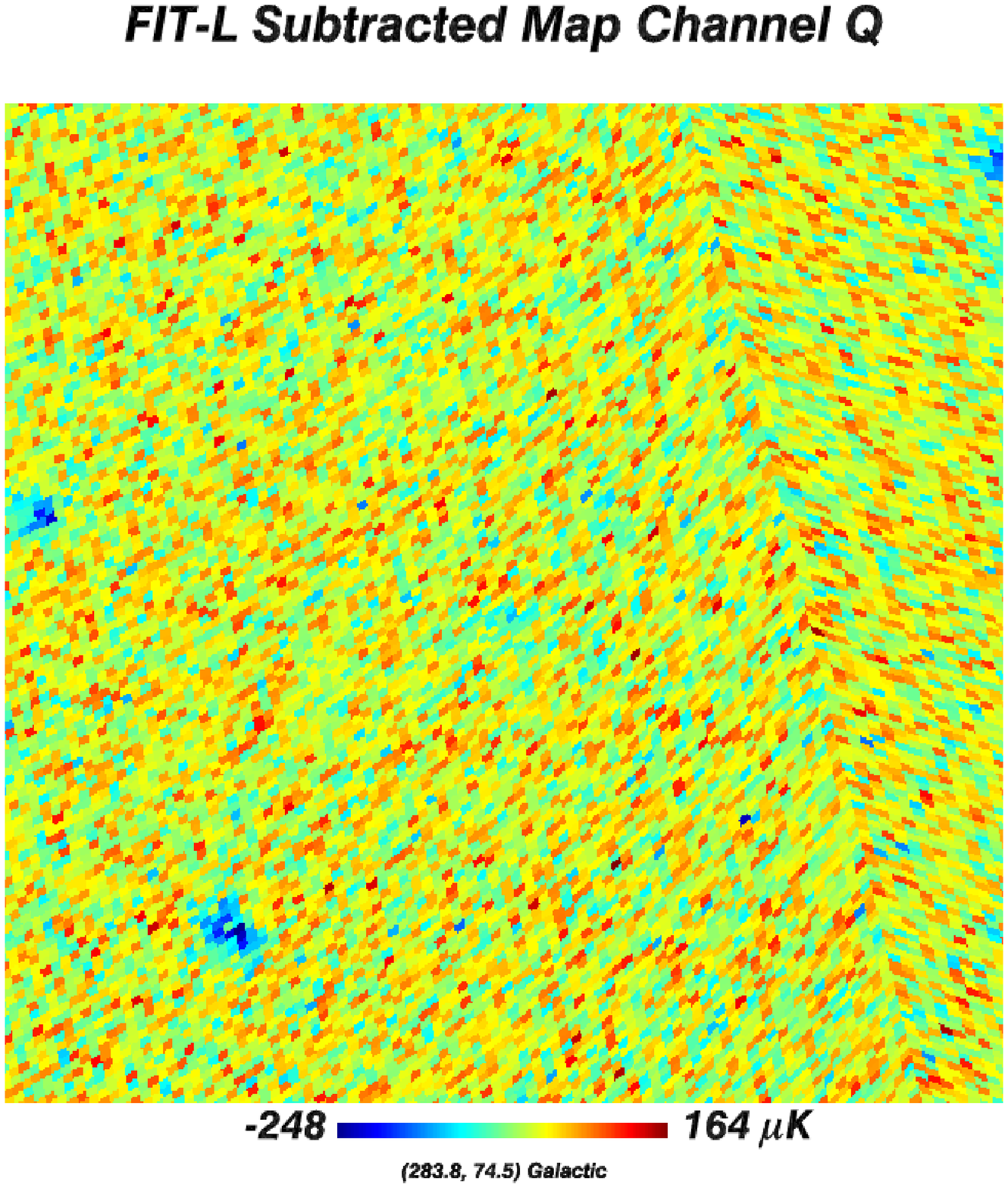} &  \includegraphics[width=4.8cm]{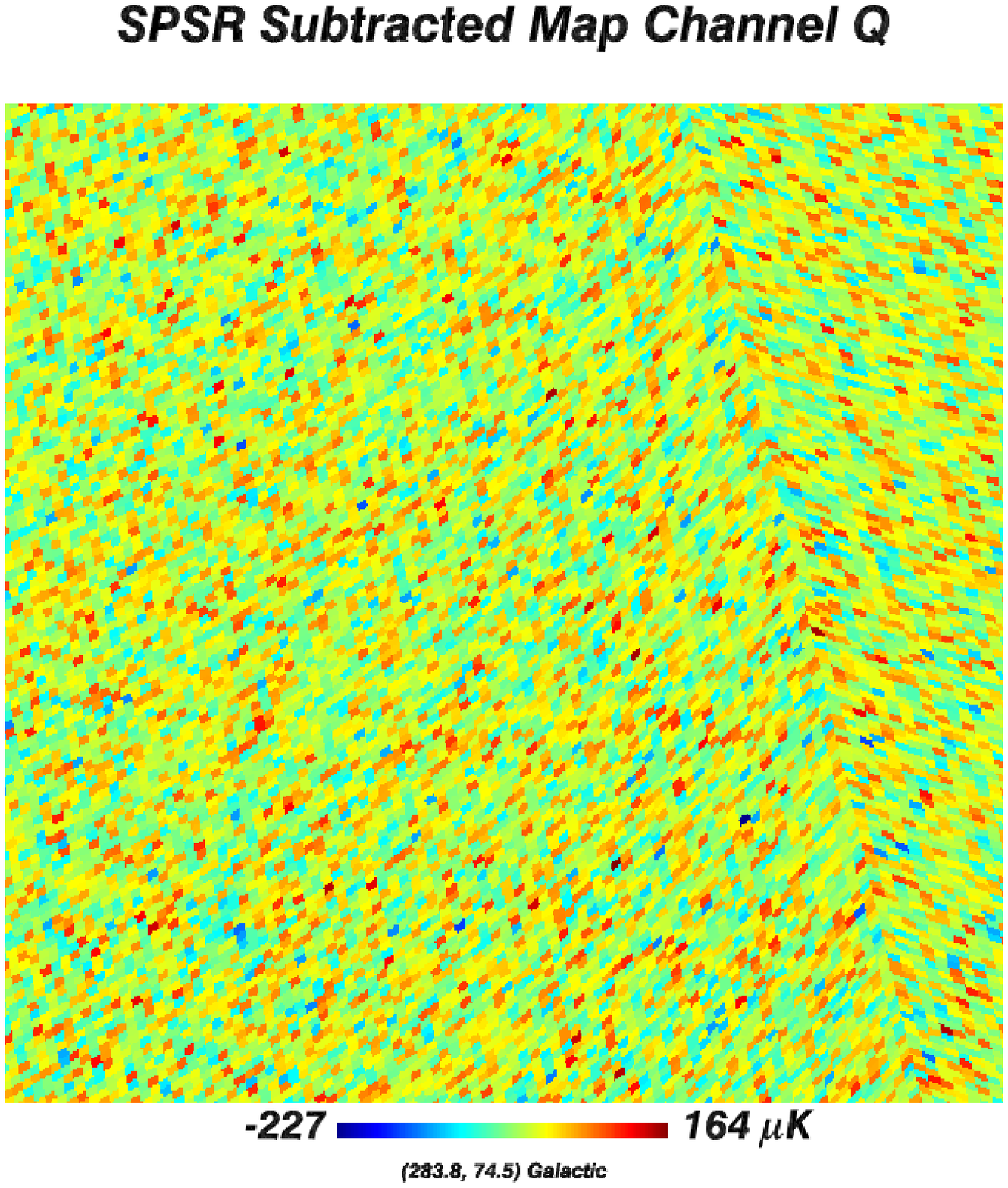}
   \end{tabular}
   \end{center}
   \caption[example] 
   { \label{fig:regionbias} Point-source residuals with noise for the various approaches in a projected region of channel Q,  illustrating the lower bias obtained for this  channel with the proposed approach. These maps were  obtained by subtracting the diffuse component from the point-source-subtracted map. In the first two maps, three negative regions can be visually detected because of the overfitting of the point-source flux in the FIT-C and FIT-L approaches, which is different from the SPSR residual map. This does not display biases that are significantly higher than the noise level.}
   \end{figure*}

Statistics on the distribution of the error in the different flux bands are displayed in Figure~\ref{fig:Boxplots}. In these box-plots both quartiles (three horizontal bars making the "box") and extreme values (upper and lower horizontal bars) are represented  for the various approaches and the different channels. First, the same conclusions can be drawn as in Table \ref{tab:pointcatfluxcut} by focusing on the interquartile range and median for the different channels: SPSR outperforms FIT-C and FIT-L, except for channel W, where results are only slightly degraded. In particular, Figure~\ref{fig:regionbias} illustrates that the bias in the lowest frequency channels is higher than the noise level in the residual maps for FIT-C and FIT-L, but not for SPSR. 

\begin{figure*}[htb]
   \begin{center}
   \begin{tabular}{ccc}
   \includegraphics[width=4.8cm]{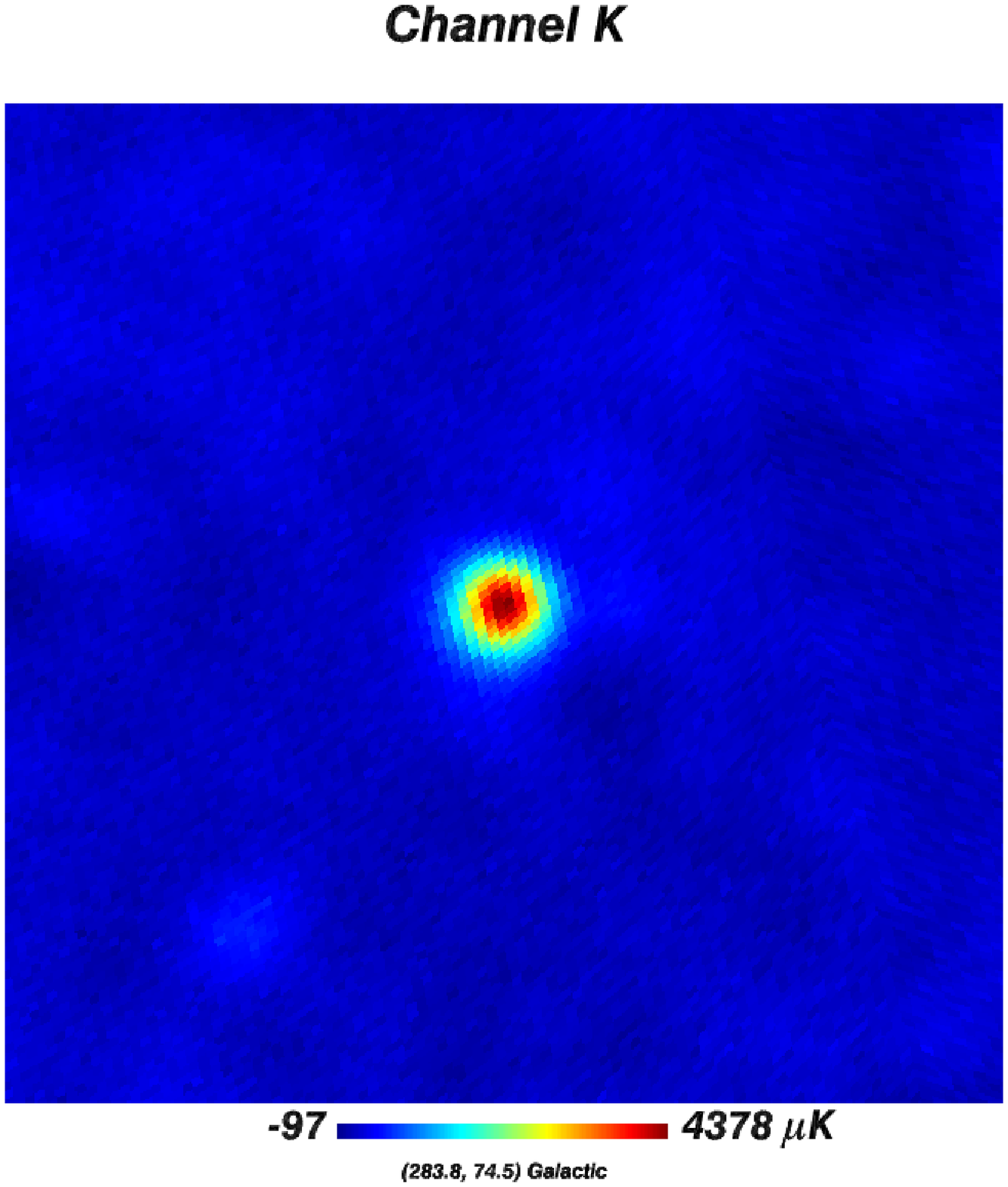} &   \includegraphics[width=4.8cm]{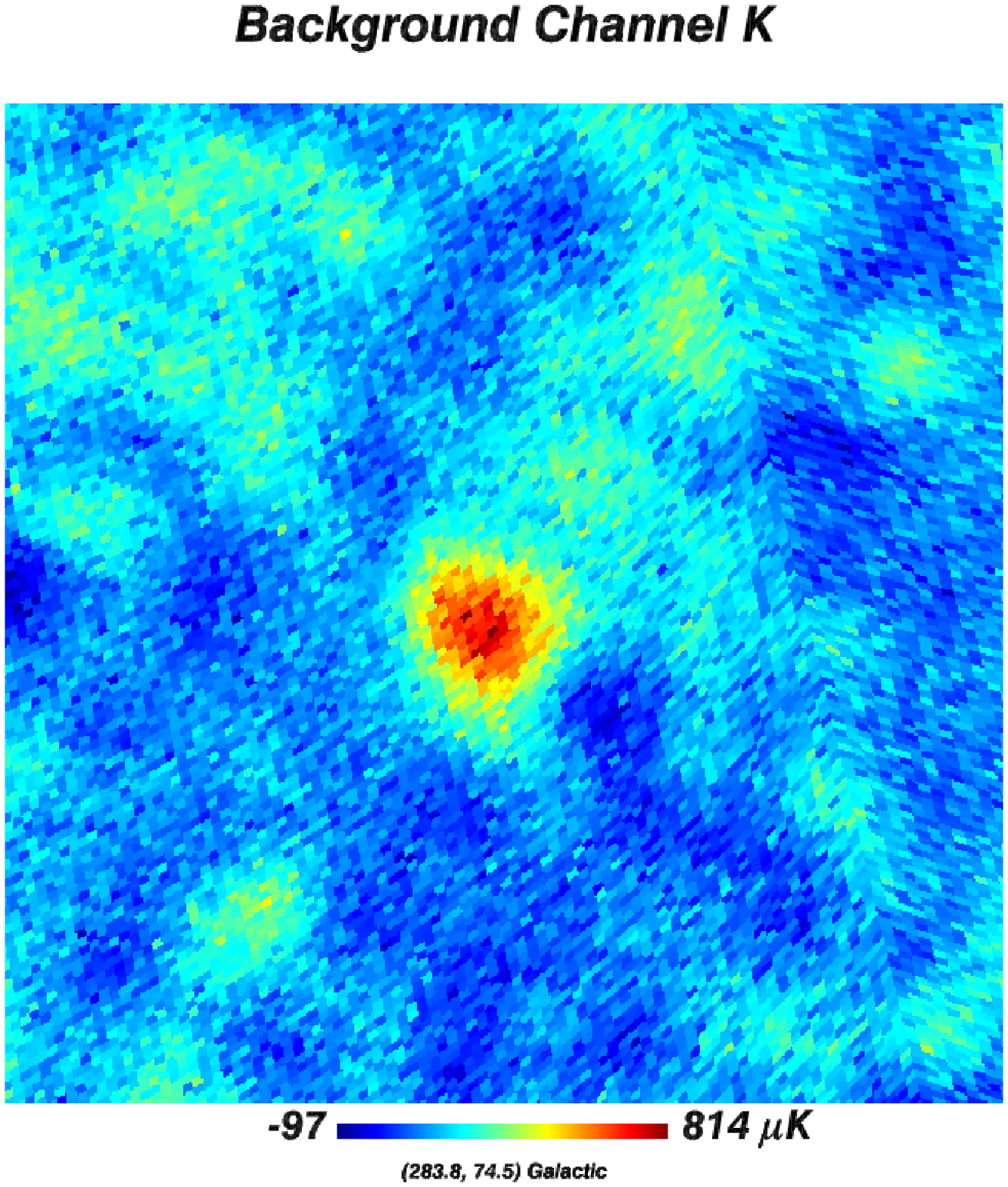} &  \includegraphics[width=4.8cm]{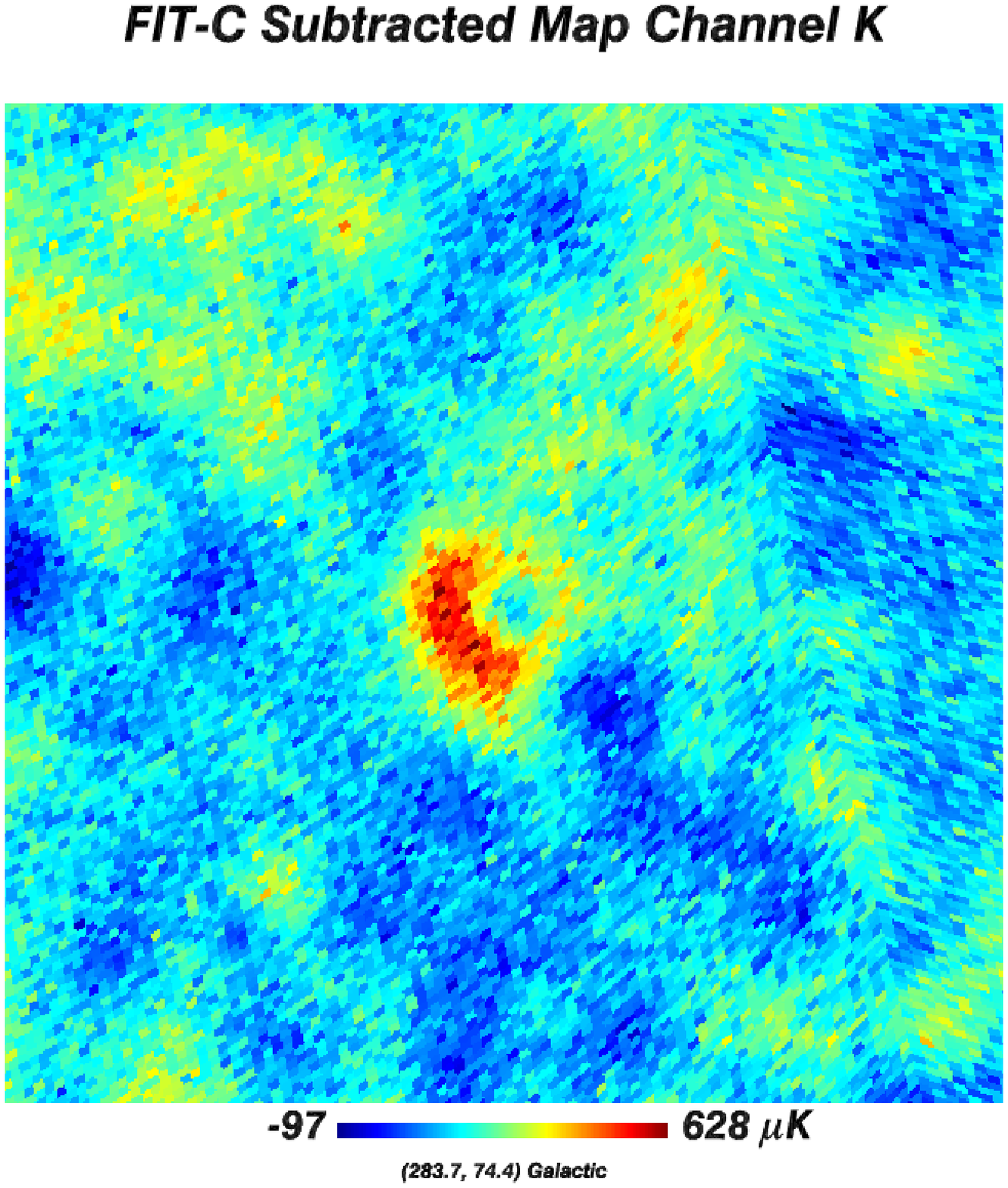}
   \end{tabular}
   \begin{tabular}{cc}
   \includegraphics[width=4.8cm]{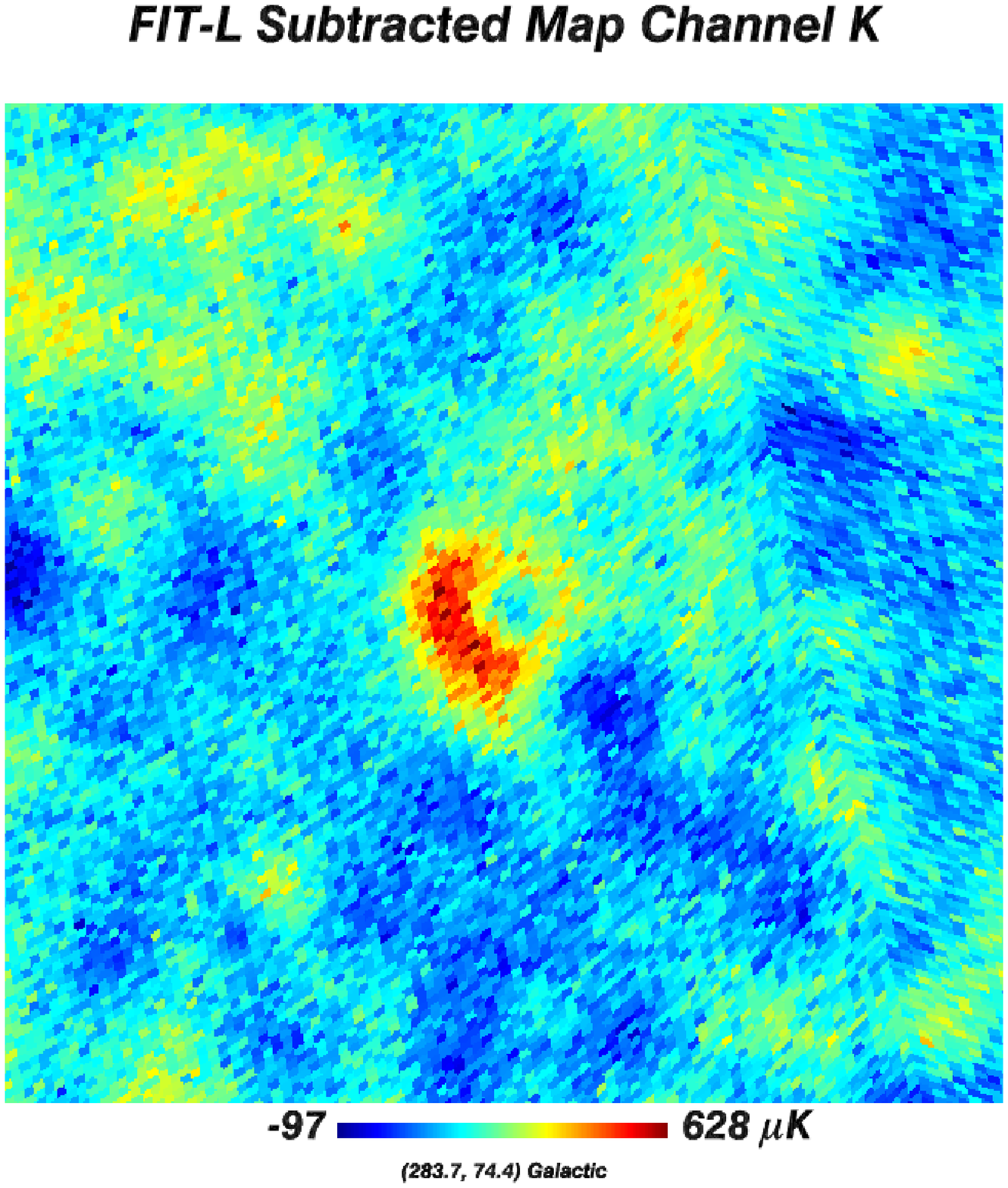} &   \includegraphics[width=4.8cm]{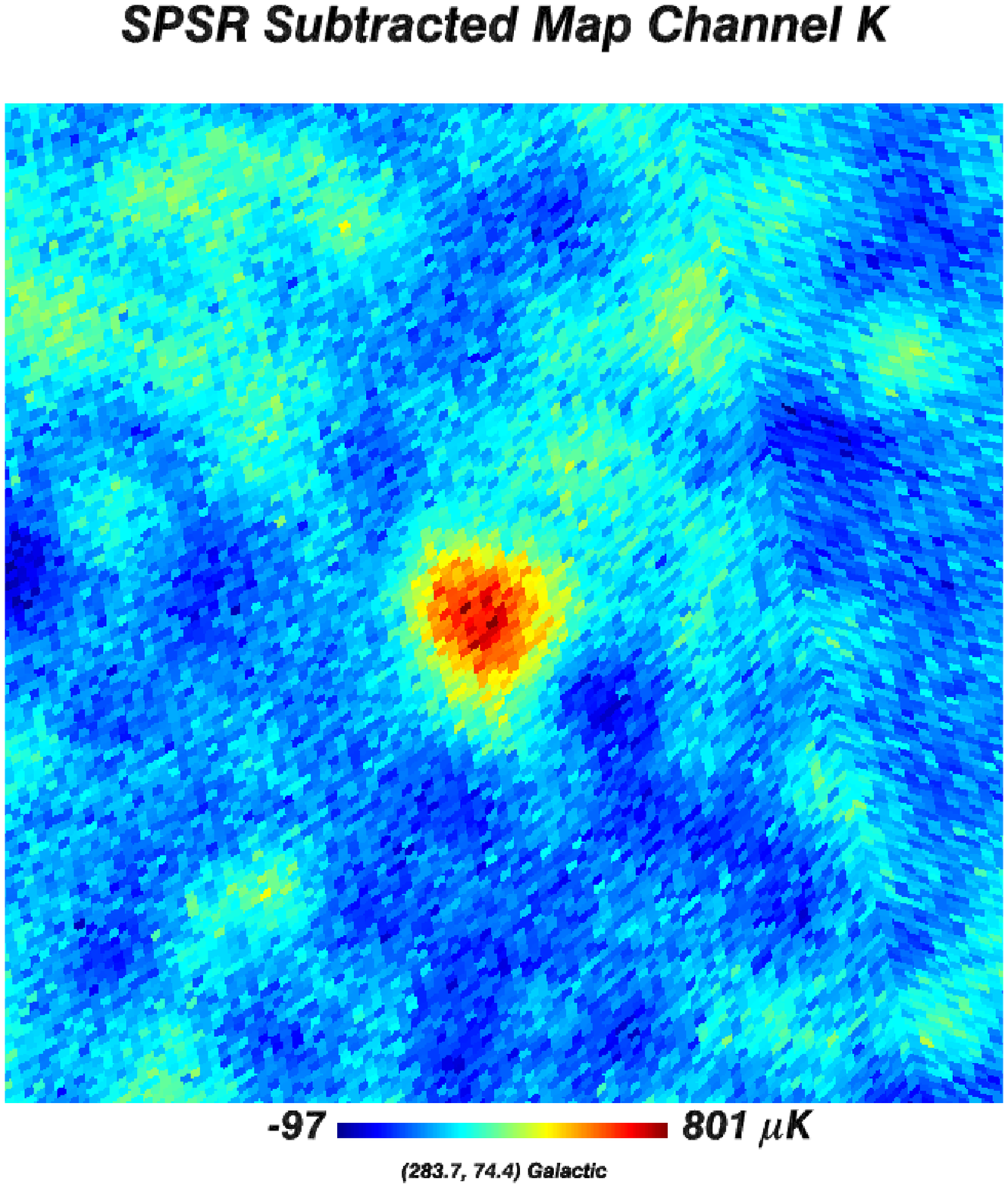}
   \end{tabular}
   \end{center}
   \caption[example] 
   { \label{fig:anomaly} Projected regions with large discrepancy between the proposed approach and the local low-order polynomial background modelling. Images correspond to the simulated data, the diffuse component with noise, and the point-source-subtracted data with the three approaches. SPSR in this case allows a better subtraction of the point-source contribution because the local variations in the background are not captured by the other two approaches.}
   \end{figure*}

These box-plots also illustrate that the proposed approach leads to lower extreme errors, while FIT-L or FIT-C can both fail with large flux errors. An example of such an incorrect flux-fitting is illustrated in Figure~\ref{fig:anomaly} for channel K, where FIT-C and FIT-L lead to overestimated flux in a complex background region that cannot be accurately modelled with low-order polynomials. In contrast, SPSR successfully estimates the background and therefore gives a better flux estimate.

\subsection{Flux recovery with internal multichannel catalogue derived from simulations}

The noise-free catalogue is useful to assess whether the more flexible background model in SPSR indeed better captures the background fluctuation than low-order local polynomials, in particular, to recover low-flux sources. However,  in practice detection is performed internally for each channel, and a catalogue is built by combining these multichannel detections. The flux-recovery problem is slightly different in this case, since the catalogue is first subject to Eddington bias (excess detection in positive fluctuations of the background, which translates into higher biases in estimating the flux and impacts low-flux sources), and then the task is also to recover as accurately as possible the spectra of sources that can be lower than the detection level in some of the channels (e.g. in higher frequencies for radio sources).

\subsubsection{Catalogue generation}

As was previously done for the \emph{WMAP} official point-source catalogue, we therefore constructed a catalogue by first applying a matched filter to each channel, taking into account second-order statistics of CMB and noise in spherical harmonic space \citep{WMAP9,Tegmark98}. Sources were considered detected if the value in the filtered map was higher than $5\sigma$ of the filtered background (computed locally). This resulted in 689 (442, 415, 248, and 172) sources for channel K (Ka, Q, V, and W). Compared with the results obtained in Table~\ref{tab:pointcatfluxcut}, a sub-population of the previous catalogue  is essentially considered at this step: sources with sufficiently high local contrast to the diffuse background that they can be internally detected. 
After merging of the sources, the final simulation-derived multichannel catalogue contained 724 sources, and we processed these in the
same way as desribed in the previous paragraph. However, after merging, the sources considered for a channel with this approach are no longer a sub-population of the previous catalogue: sources detected in other bands are considered as well.

\subsubsection{Flux recovery and point-source subtraction}

The same working conditions were set for SPSR as in the previous simulations. The statistics obtained as in Table~\ref{tab:RMSE_fluxcut} are presented in Table~\ref{tab:RMSE_fluxcut_detect}. The values reported for medium- or high-flux point sources in the first channels are similar if not identical, because  this sub-population did not change from the previous catalogue. However, significant discrepancies from Table~\ref{tab:RMSE_fluxcut}  can be noted. First, the biases have increased for all methods for low-flux sources, an indication for Eddington bias. Note, however, that the bias has increased less for SPSR than for FIT-C or FIT-L in absolute values, leading to even larger differences between the methods (SPSR leads to the lowest biases for all channels, with a bias now up to more than $100\textrm{mJy}$ lower for sources lower than $1\textrm{Jy}$ in channel K). This is also true for the RMSE of these low-flux sources, which essentially did not change for SPSR and was slightly degraded for FIT-C or FIT-L. 
Note that most of the low-flux sources in the catalogue are not detected in the W band, which might explain why SPSR gave better RMSE for medium-flux sources than FIT-C or FIT-L, while this was not the case in the noise-limited catalogue. The proposed approach in this multichannel scenario therefore leads to the lowest errors, even for flux estimated in channel W. 

\begin{figure*}[htb]
  \begin{center}
  \begin{tabular}{ccc}
   \includegraphics[width=4.8cm]{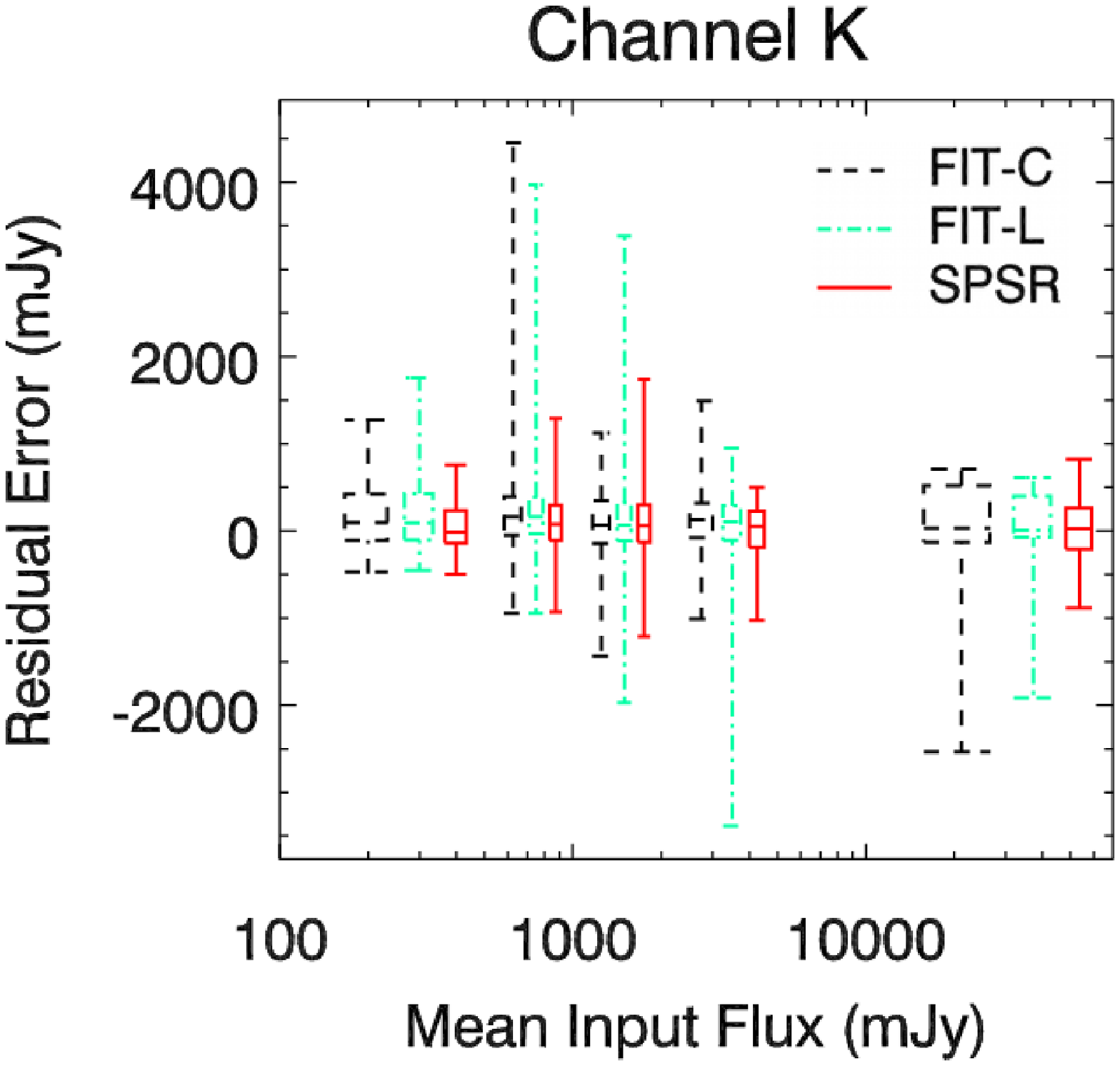} &   \includegraphics[width=4.8cm]{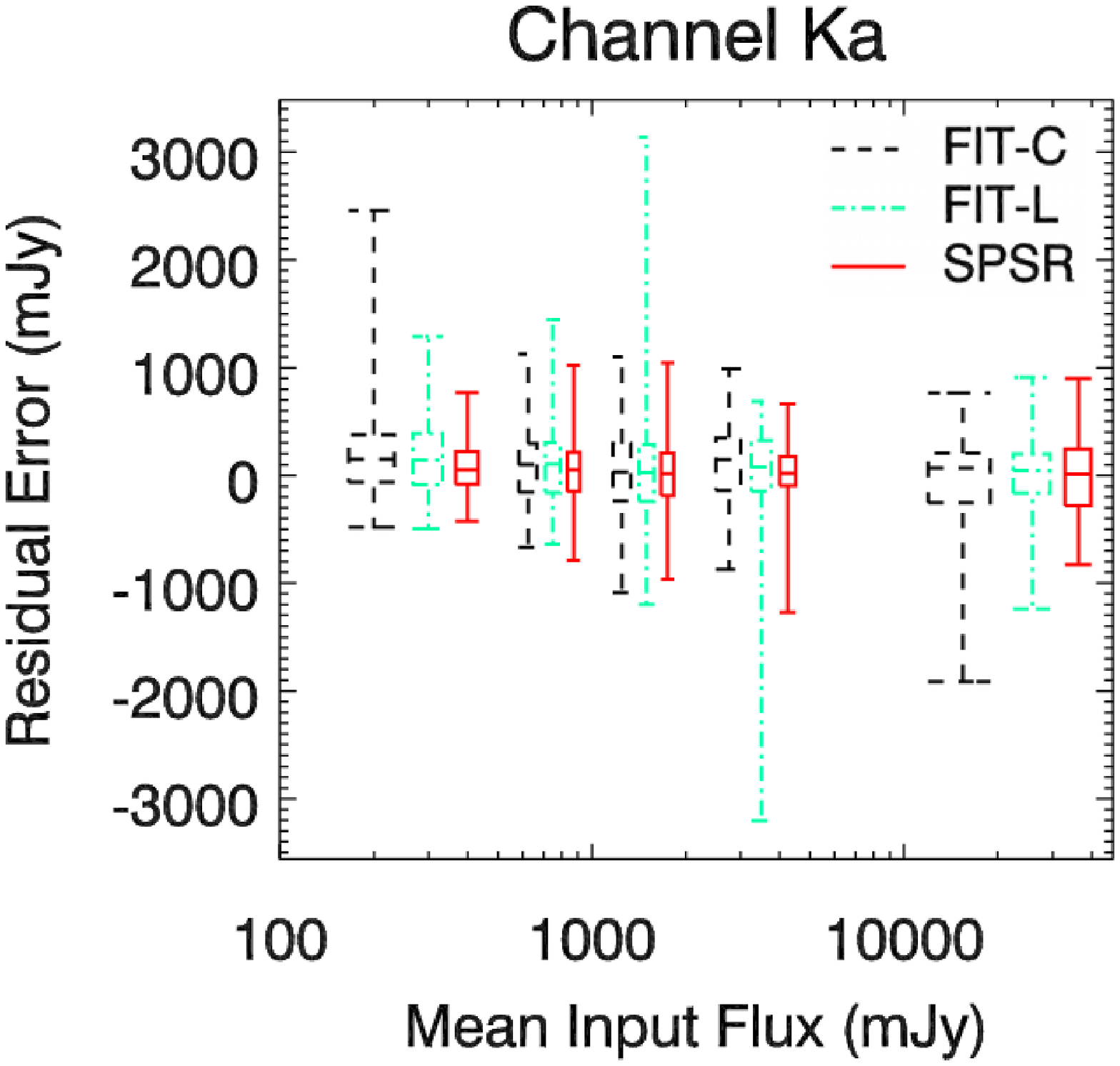} & \includegraphics[width=4.8cm]{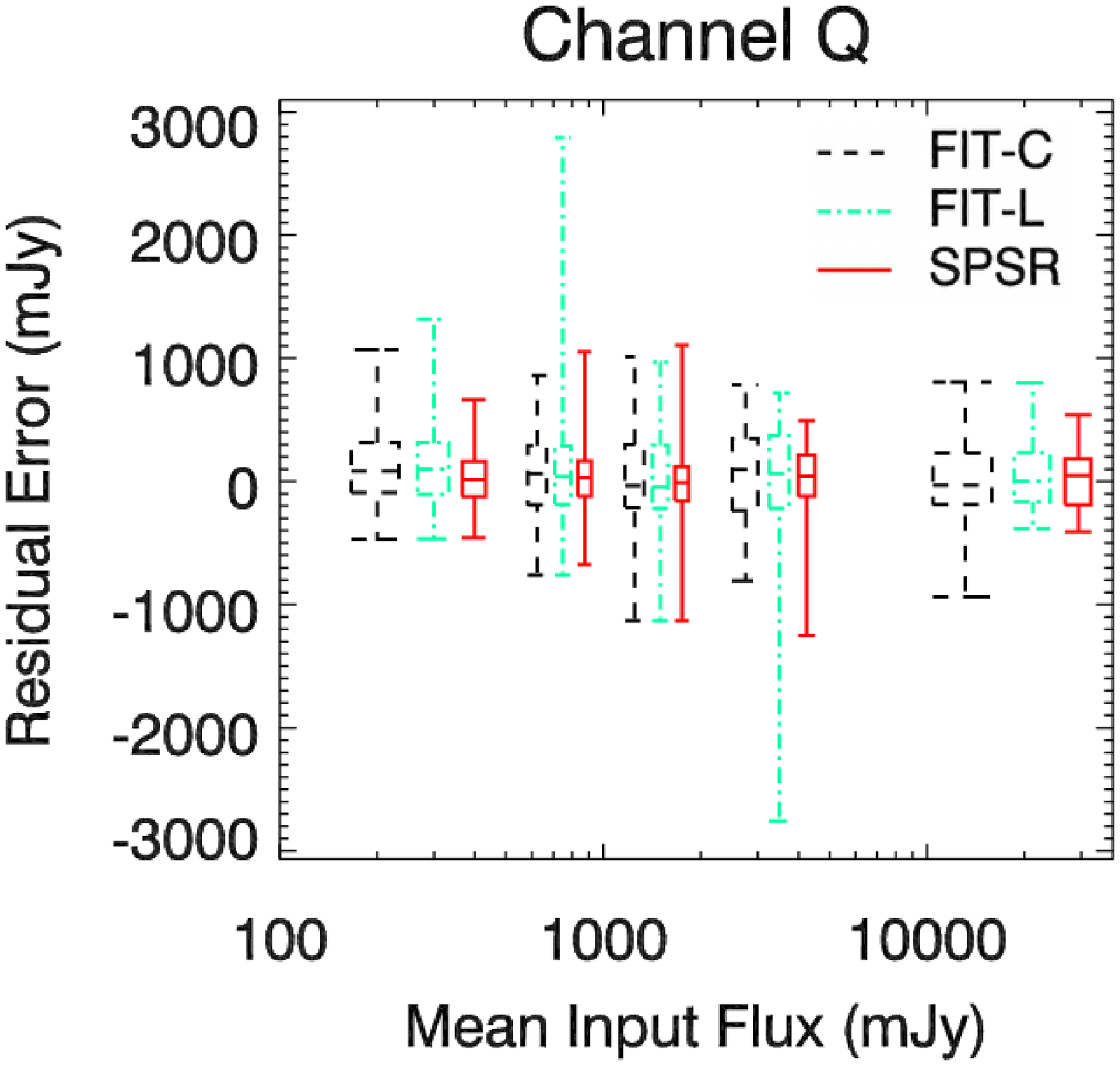}
   \end{tabular}
   \begin{tabular}{cc}
     \includegraphics[width=4.8cm]{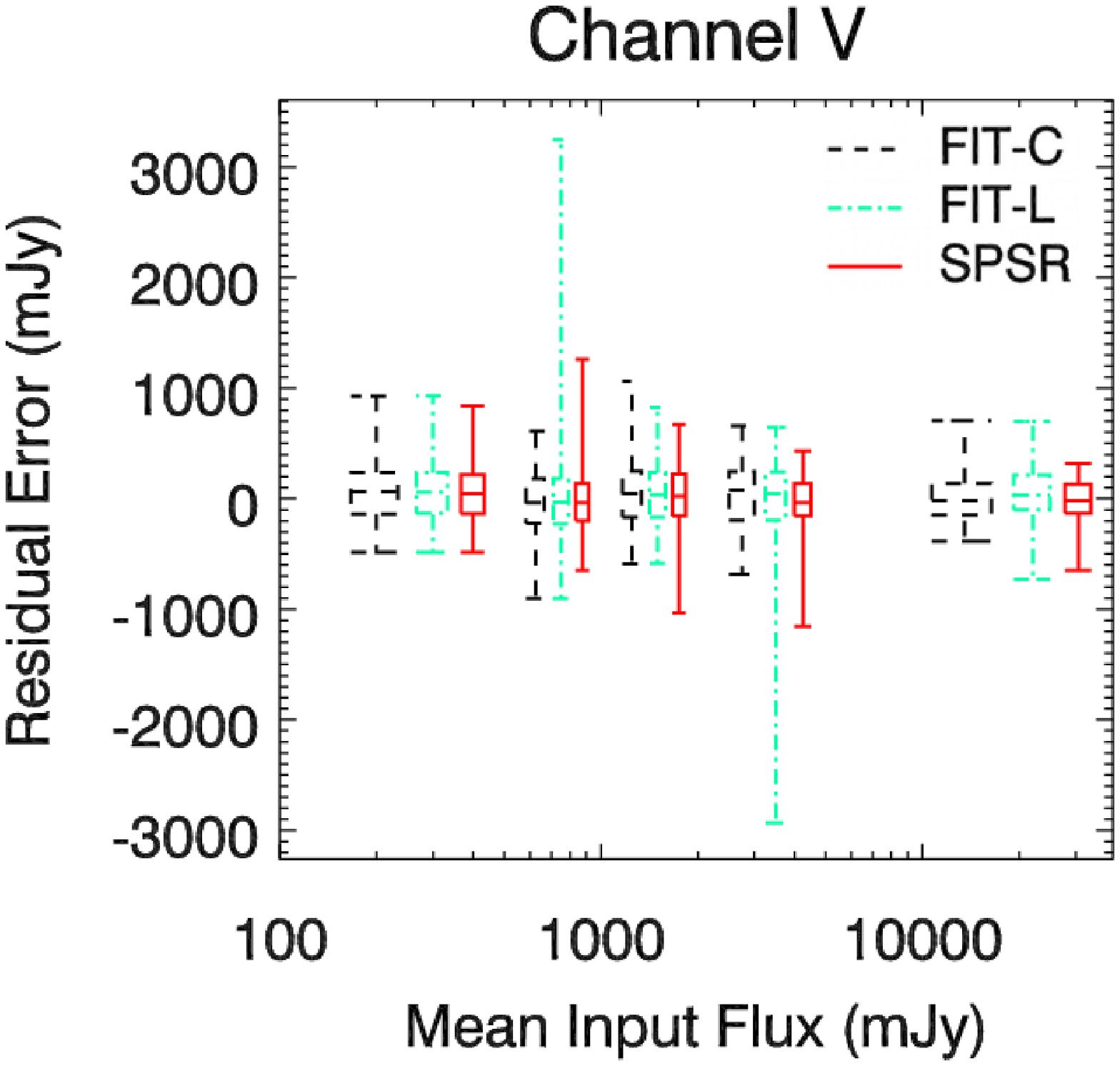} &   \includegraphics[width=4.8cm]{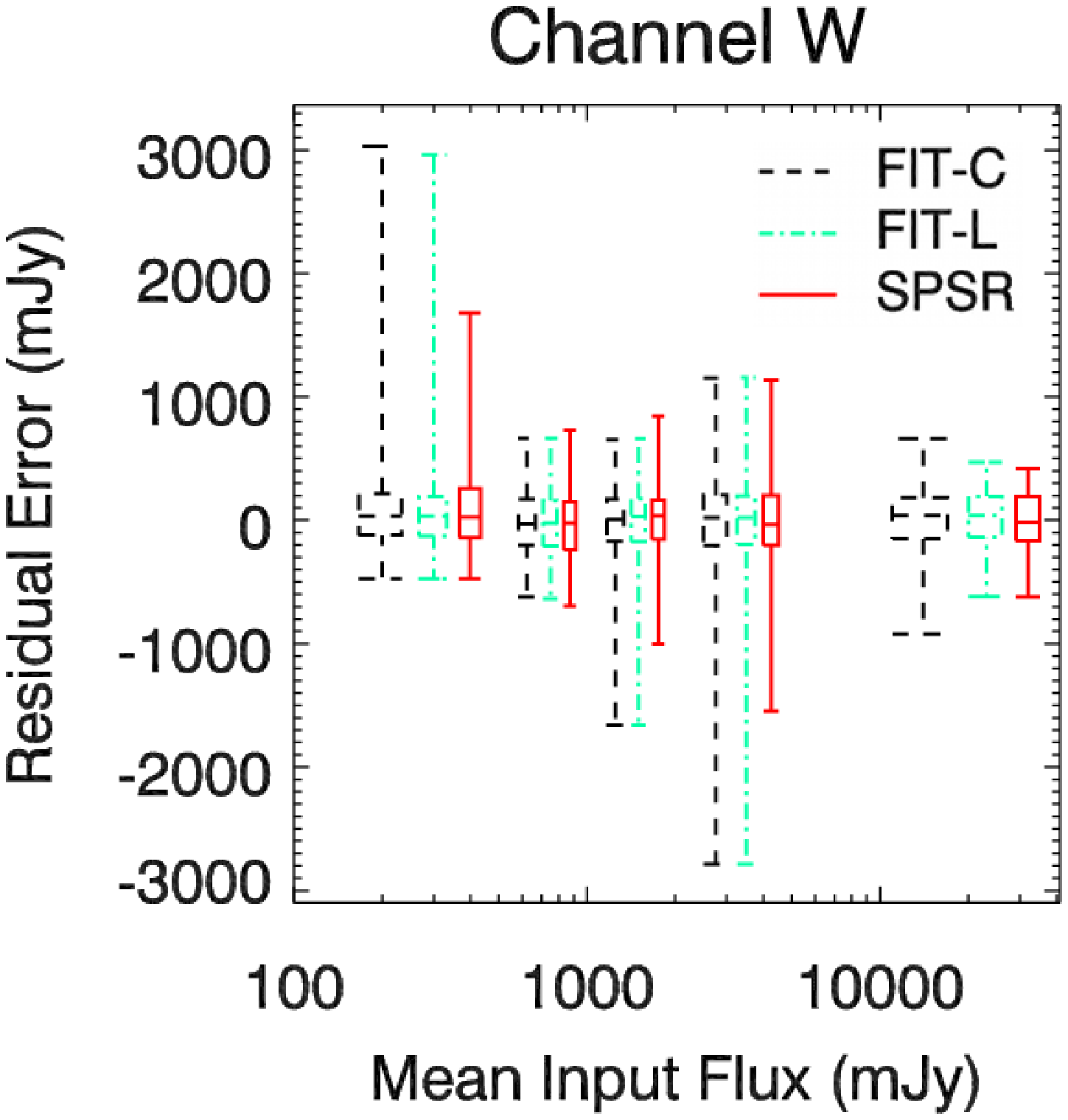}\\
   \end{tabular}
   \end{center}
  \caption[example] 
   { \label{fig:Boxplots_detect} Statistics on error flux computed from internal catalogue and using the various approaches for the five channels  for five different flux bands (flux $< 500$ mJy, $500$ mJy $\leq $ flux $< 1$ Jy, $1$ mJy $\leq $ flux $< 2$ Jy, $2$ Jy $\leq $ flux $< 5$ Jy, flux $\geq 5$ Jy).  Quartiles and extreme values are plotted. }
   \end{figure*}

The distribution of errors is reported in Figure~\ref{fig:Boxplots_detect}. SPSR seems as robust to large flux errors in the internal catalogue as in the noise-limited case with respect to FIT-C or FIT-L.  Note the distribution of errors for channel W for low-flux sources (lower than $500\textrm{mJy}$), which were indeed not detected in this channel (compare with Figure~\ref{fig:Boxplots}). Large flux errors are observed for all methods, but are less pronounced  for SPSR than for FIT-C or FIT-L.

As a summary for the internal catalogue, the most striking difference compared with the noise-limited catalogue occurs in low-flux sources that are now subject to Eddington bias, whereas for high-flux the results are essentially similar. Better relative performance in this situation in terms of bias and RMSE or MAD may be attributed to two different phenomena: the proposed approach would be less affected by Eddington bias and/or lead to better estimates for sources not detected in the band.

\begin{table*}[htb]
\caption{Statistics on the recovered flux for each channel using the multichannel internal catalogue for the three approaches and for various flux bands (in mJy). Again, the best results are displayed in bold.} 
\label{tab:RMSE_fluxcut_detect}
\begin{center}       
\begin{tabular}{|l|l|l|l|l|l|l|l|l|l|l|} 
\hline
\rule[-1ex]{0pt}{3.5ex} & & \multicolumn{3}{c|}{Flux $<$ 1 Jy} & \multicolumn{3}{c|}{1 Jy $\leq$ Flux $<$ 5 Jy} & \multicolumn{3}{c|}{Flux $\geq$ 5 Jy} \\
\rule[-1ex]{0pt}{3.5ex} Channel &Method  & Bias & RMSE & MAD & Bias & RMSE & MAD & Bias & RMSE &MAD \\
\hline
\rule[-1ex]{0pt}{3.5ex}  \multirow{4}{*}{K} & FIT-C & 171.3 & 437.7 & 311.7 & 112.8 & 385.1 & 292.9 & 68.4  & 650.6 &   404.2\\
\rule[-1ex]{0pt}{3.5ex} & FIT-L & 168.5 & 422.6 & 305.3 & 97.5  & 472.6 & 300.1 & 90.0 & 518.5 &   317.5\\
\rule[-1ex]{0pt}{3.5ex}  & SPSR & \tb{70.3} & \tb{337.6}& \tb{257.7} & \tb{46.6} & \tb{364.3} & \tb{274.1} & \tb{0.4} & \tb{380.1}  &  \tb{275.0} \\
\hline
\rule[-1ex]{0pt}{3.5ex}  \multirow{4}{*}{Ka} & FIT-C & 115.5 & 381.0 & 296.0 & 50.4 & 392.6 & 312.7 & -26.9  & 548.1 &   356.3\\
\rule[-1ex]{0pt}{3.5ex} & FIT-L & 115.5 & 371.7 & 296.7 & 38.7  & 468.4 & 313.7 & 30.6 & 508.5 &   378.9\\
\rule[-1ex]{0pt}{3.5ex}  & SPSR & \tb{49.3} & \tb{268.2} & \tb{210.7} & \tb{6.6} & \tb{293.9} & \tb{221.9} & \tb{11.9} & \tb{376.0}  &  \tb{ 289.2} \\
\hline
\rule[-1ex]{0pt}{3.5ex}  \multirow{4}{*}{Q} & FIT-C & 75.4 & 334.1 & 267.1 & 30.0 & 355.9 & 284.4 & -26.2  & 373.7 &   280.3\\
\rule[-1ex]{0pt}{3.5ex} & FIT-L & 79.1 & 354.7 & 269.5 & 11.5  & 400.6 & 289.3 & 22.6 & 305.6 &   239.6\\
\rule[-1ex]{0pt}{3.5ex}  & SPSR & \tb{30.4} & \tb{227.2} & \tb{177.1} & \tb{6.9} & \tb{255.5} & \tb{185.1} & \tb{22.4} & \tb{236.9}  &   \tb{191.5} \\
\hline
\rule[-1ex]{0pt}{3.5ex}  \multirow{4}{*}{V} & FIT-C & 27.3 & 286.4 & 228.4 & 37.9 & 300.0 & 243.5 & \tb{21.1}  & 248.2 &   178.0\\
\rule[-1ex]{0pt}{3.5ex} & FIT-L & 32.8 & 314.8 & 230.8 & 16.3  & 353.7 & 245.3 & 53.5 & 312.8 &   224.0\\
\rule[-1ex]{0pt}{3.5ex}  & SPSR & \tb{17.6} & \tb{259.6} & \tb{202.8} &\tb{7.1} & \tb{274.0} & \tb{211.8} & -23.3 & \tb{230.1}  &  \tb{172.8} \\
\hline
\rule[-1ex]{0pt}{3.5ex}  \multirow{4}{*}{W} & FIT-C & 33.8 & 289.1 & 201.5 & -11.0 & 357.2 & 230.0 & \tb{4.4}  & 346.1 &   254.3\\
\rule[-1ex]{0pt}{3.5ex} & FIT-L & 32.3 & 286.6 & \tb{200.7} & -10.7  & 353.8  & 227.2 & 13.7 & 285.6 &   225.4\\
\rule[-1ex]{0pt}{3.5ex}  & SPSR & \tb{30.2} & \tb{277.7} & 210.3 &  \tb{10.3} &  \tb{302.8} &  \tb{220.3} & -8.3 & \tb{253.7}  &  \tb{198.1} \\
\hline
\end{tabular}
\end{center}
\end{table*} 

\section{\label{sec:wmap9data} WMAP 9-year processing}

The proposed approach was then applied to subtract point-source emission from \emph{WMAP} 9-year data. The maps were processed as follows: the differential assemblies data deconvolved with the asymmetrical part of the beam were averaged to obtain 9-year frequency band data \citep{WMAP9}. The beam considered for each channel was obtained as the mean over all axisymetric beams provided for the differential assemblies at that frequency. A catalogue was then built by merging the two catalogues provided by the \emph{WMAP} collaboration, resulting in 628 considered point sources.

\begin{figure*}[htb]
   \begin{center}
   \begin{tabular}{ccc}
  \includegraphics[width=4.8cm]{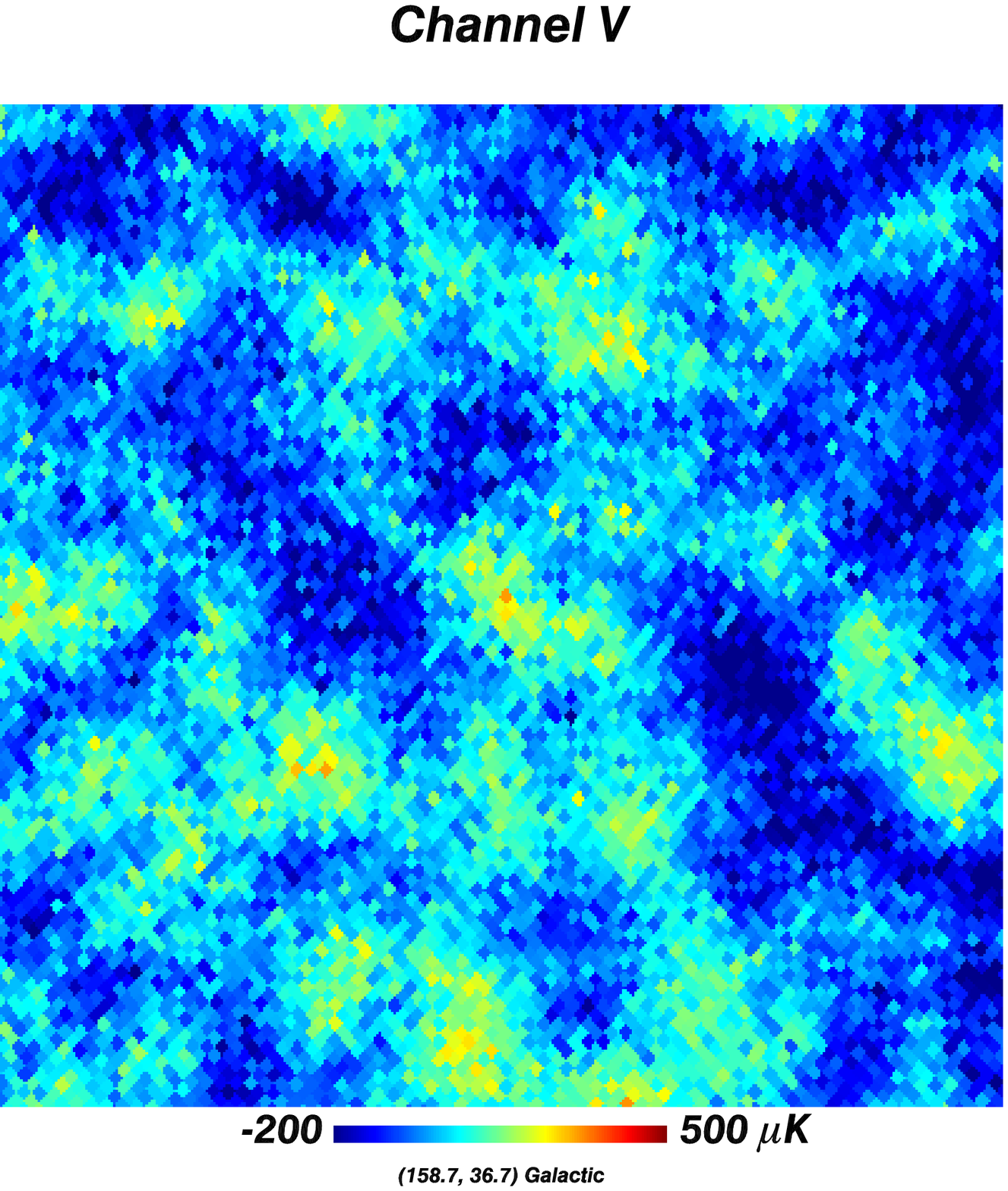} &   \includegraphics[width=4.8cm]{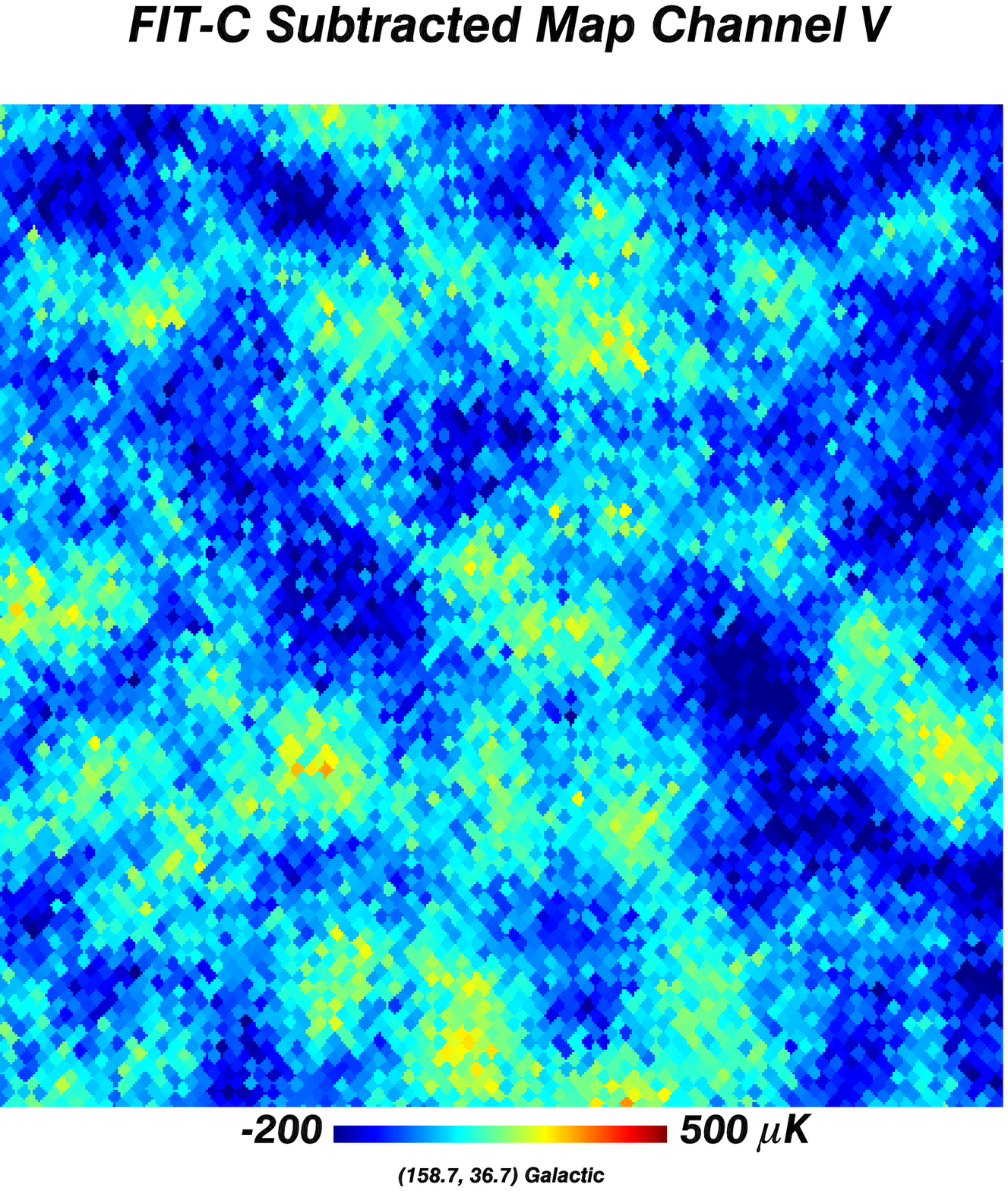} &   \includegraphics[width=4.8cm]{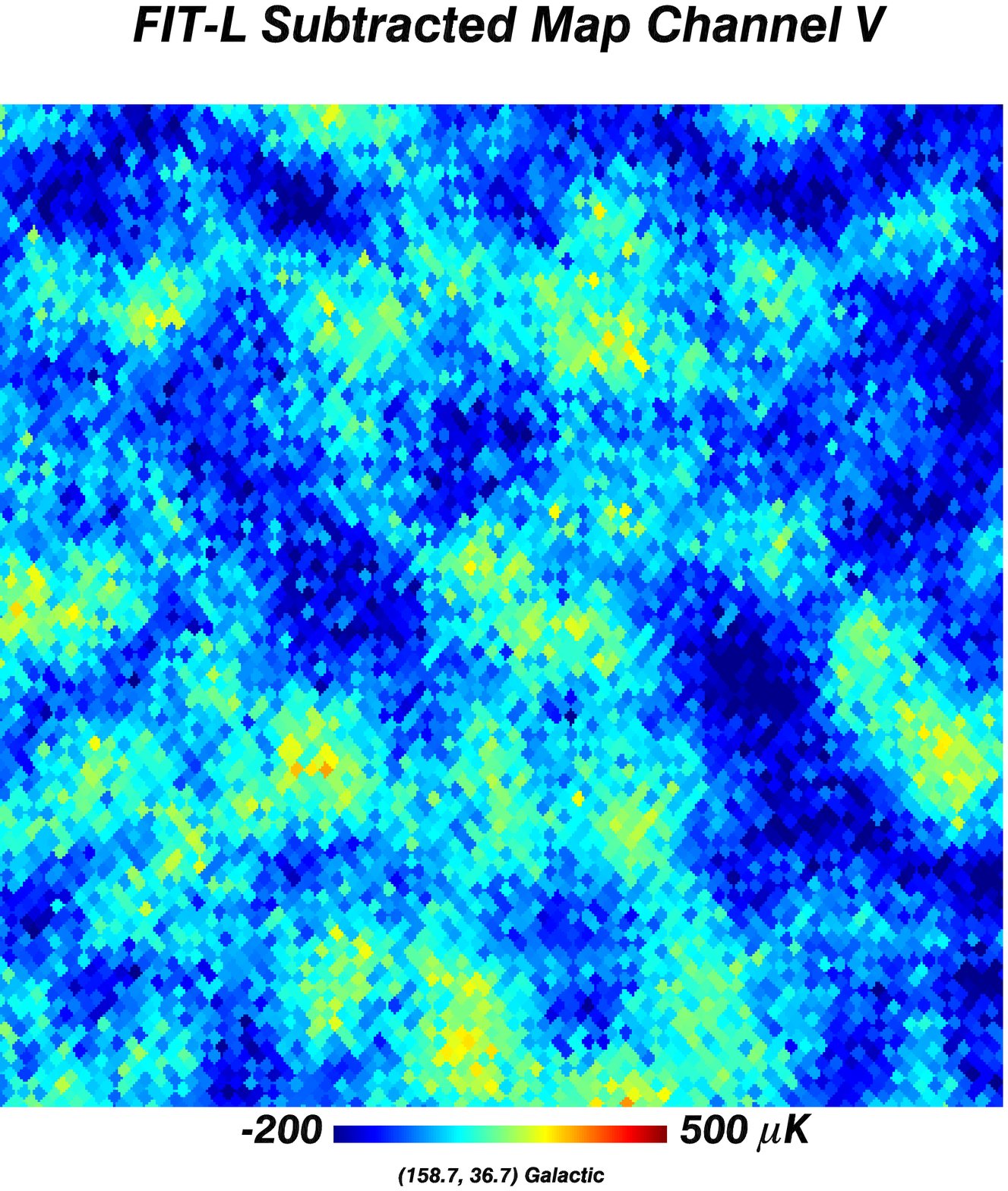} 
   \end{tabular}
   \begin{tabular}{cc}
   \includegraphics[width=4.8cm]{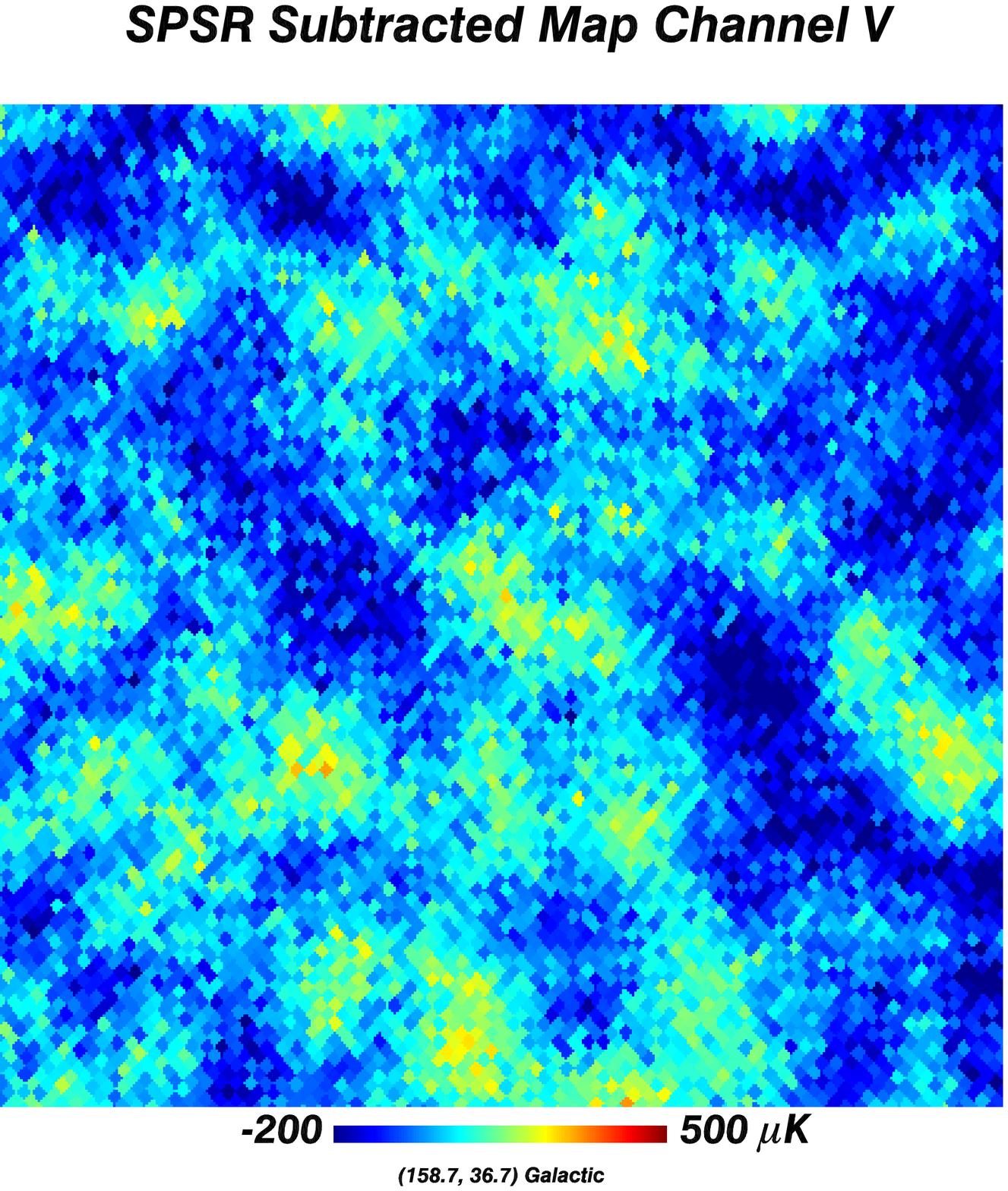} &   \includegraphics[width=4.8cm]{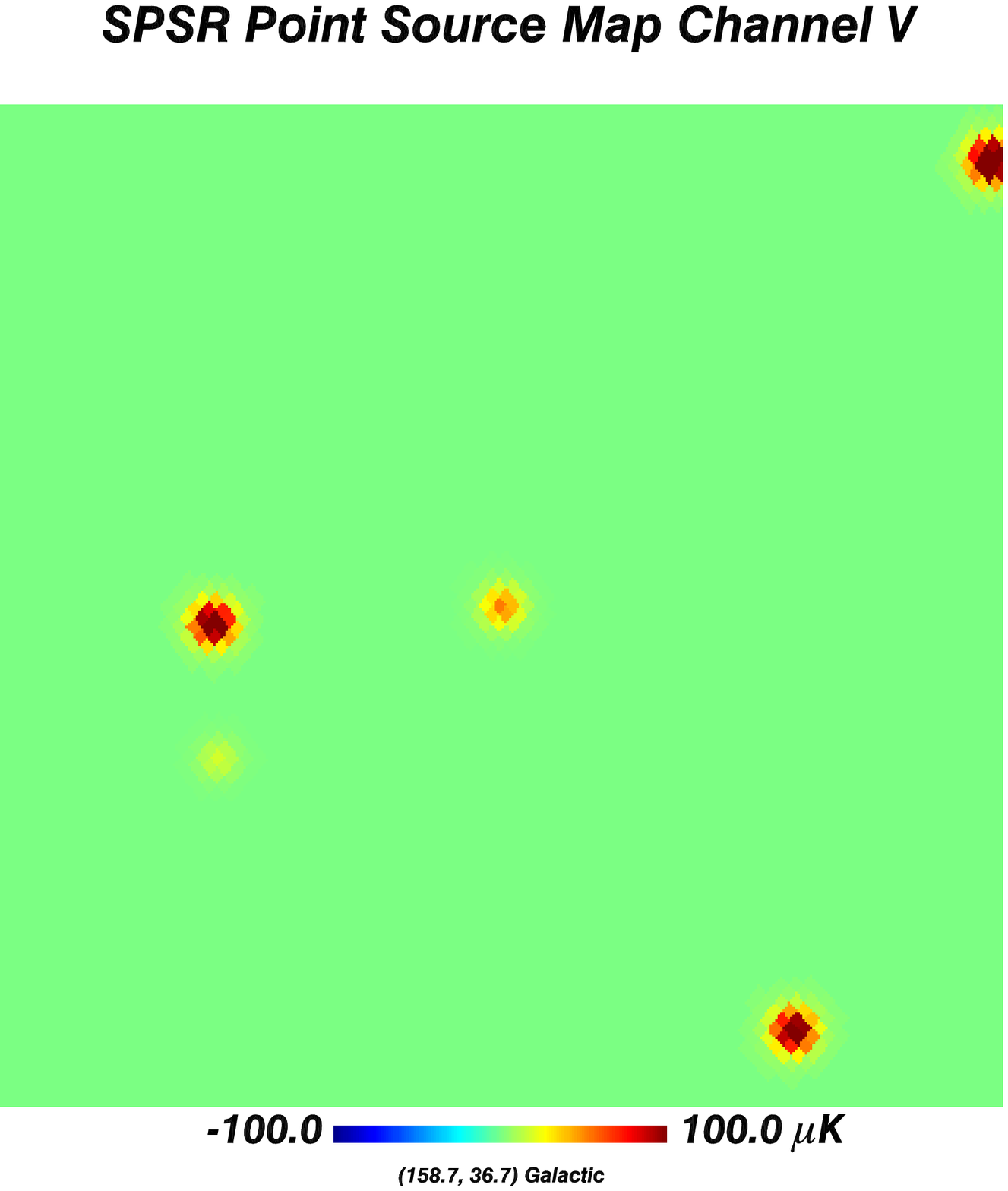}
   \end{tabular}
   \end{center}
   \caption[example] 
   { \label{fig:WMAPanomaly} Projected region in channel V of the \emph{WMAP} 9-year data showing the largest discrepancy between SPSR and FIT-C or FIT-L estimated fluxes for two fitted sources. Images correspond to the \emph{WMAP} data, the point-source-subtracted data with FIT-C, FIT-L, and SPSR approaches, and point-source-estimated data using SPSR. The differences here between the approaches are a fraction of noise and the diffuse component levels.}
   \end{figure*}

\begin{figure*}[htb]
   \begin{center}
   \begin{tabular}{cc}
   \includegraphics[width=5cm]{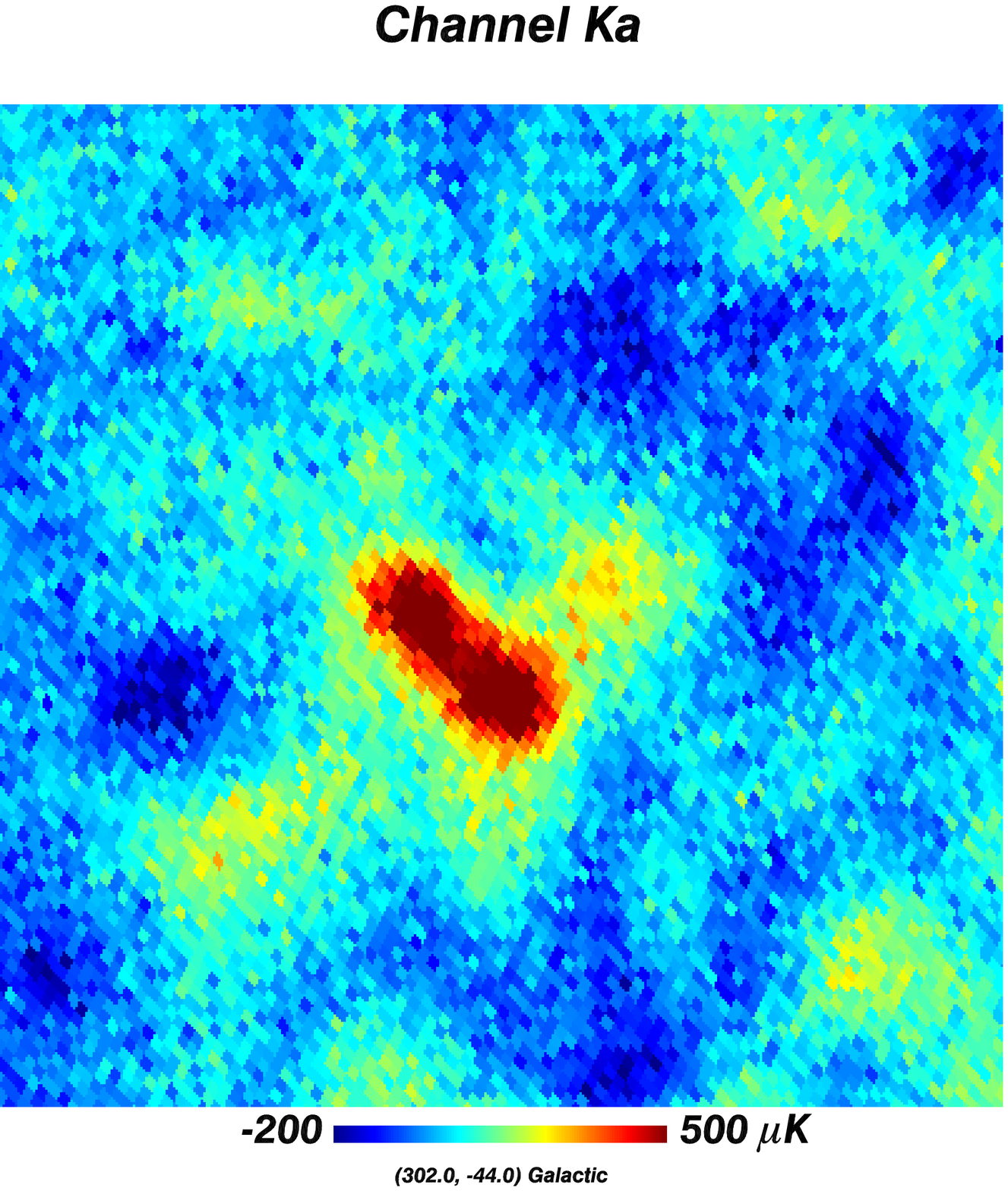} &   \includegraphics[width=5cm]{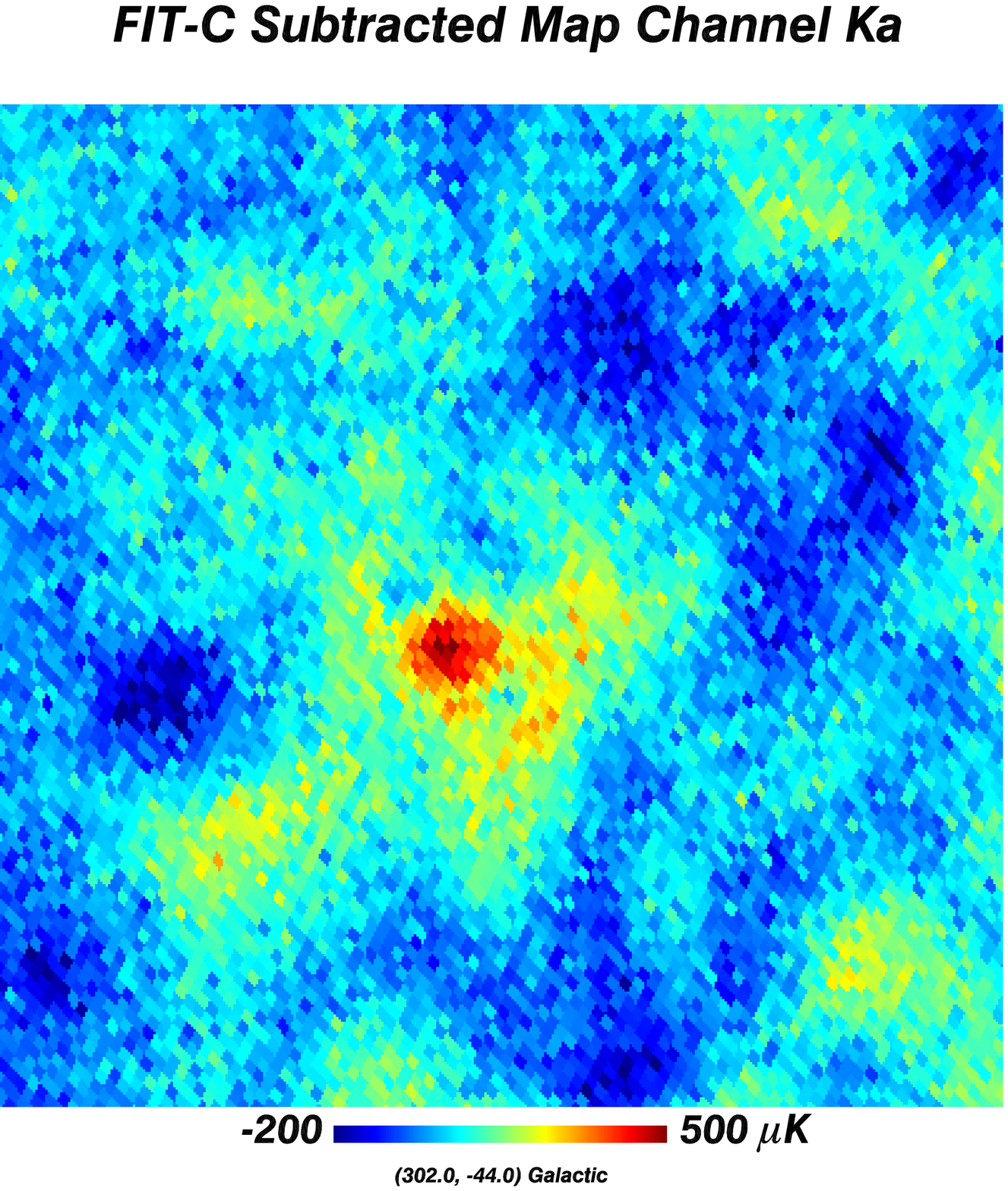} 
   \end{tabular}
   \begin{tabular}{cc}
   \includegraphics[width=5cm]{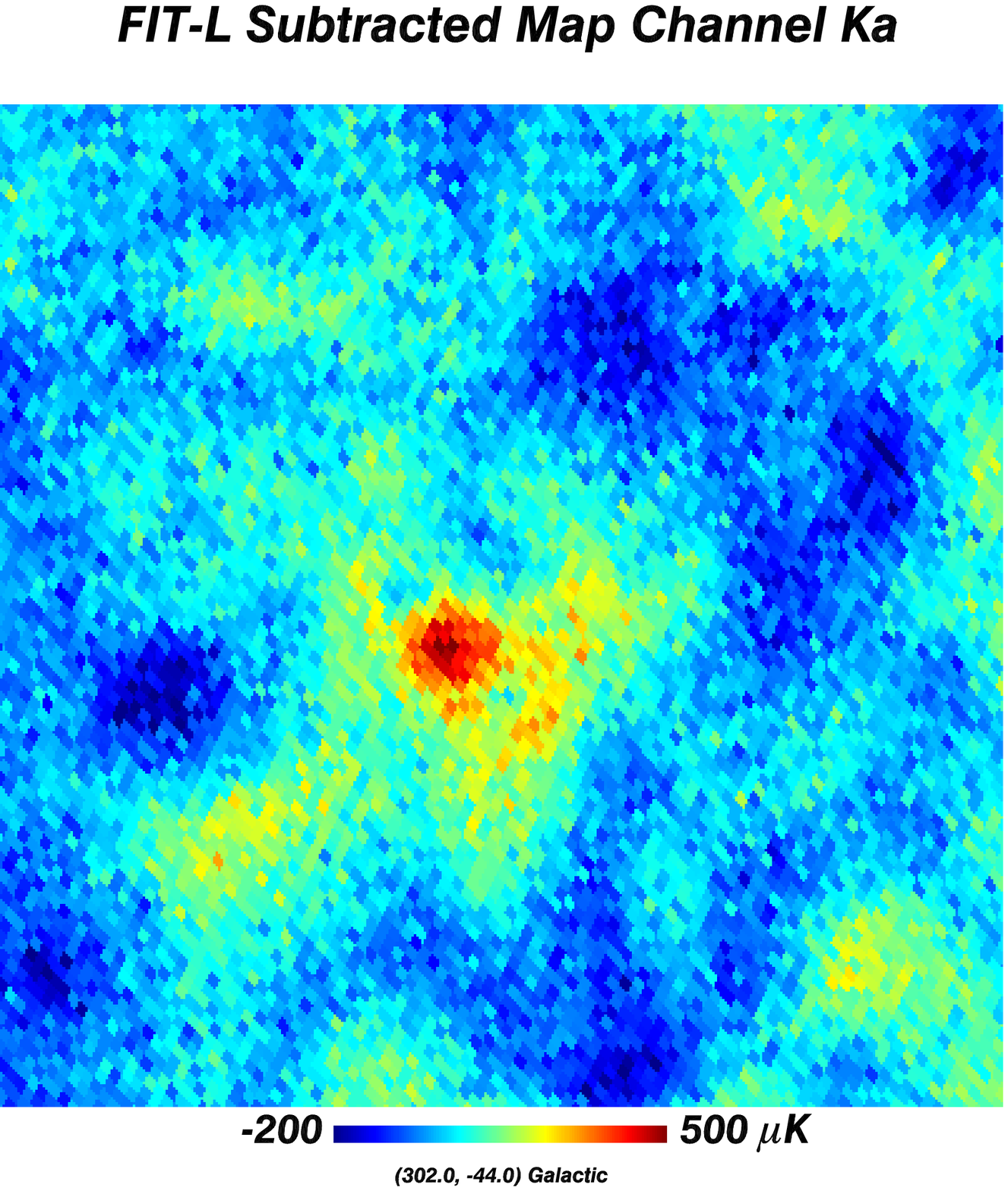} &   \includegraphics[width=5cm]{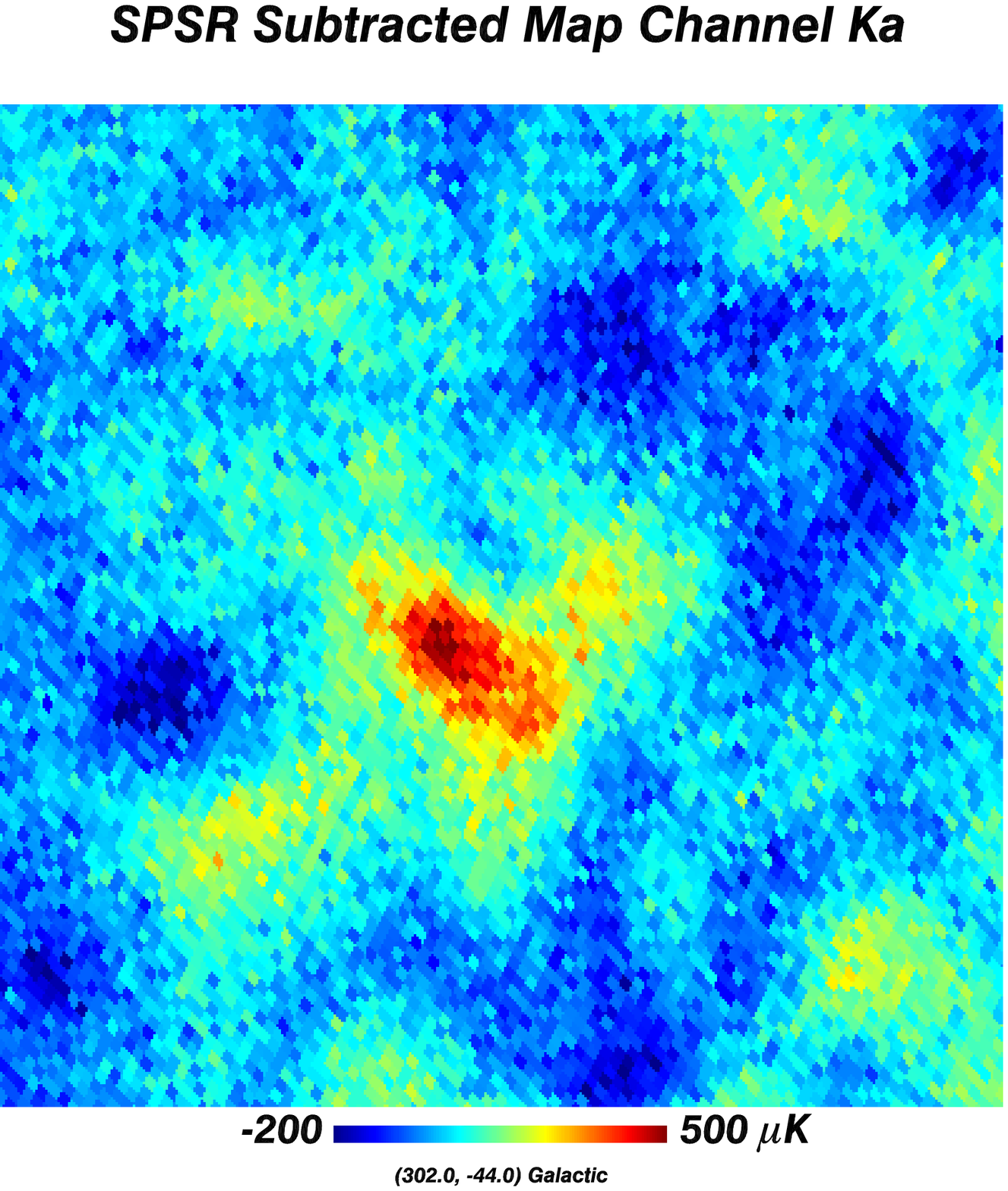}
   \end{tabular}
   \end{center}
   \caption[example] 
   { \label{fig:WMAPanomaly2} Projected region in \emph{WMAP} 9-year data showing the large discrepancy in between SPSR and FIT-C or FIT-L estimated fluxes for two fitted sources. Images correspond to the \emph{WMAP} data, the point-source-subtracted data with FIT-C, FIT-L, and SPSR approaches. The positive ring around two the fitted regions as well as a negative peak inside these regions seem to indicate that FIT-C or FIT-L are here over-fitting. The same phenomenon is absent for SPSR.}
   \end{figure*}

The same processing (with the same working conditions) as for the simulations was then performed for both SPSR and the local low-order polynomial minimization. In the absence of ground truth, it is obviously difficult to quantitatively assess the relative performance of the algorithm as in the simulation. Only extreme cases that visually illustrate the performance of the approaches are therefore presented in Figs.~\ref{fig:WMAPanomaly} and~\ref{fig:WMAPanomaly2}.

Figure~\ref{fig:WMAPanomaly} represents a patch centred on the source with the strongest difference between SPSR and FIT-L for channel V. Because the difference between the approaches is only a fraction of the background fluctuations at larger scales than the sources, visually no difference can be observed for that channel. 
However, Figure~\ref{fig:WMAPanomaly2} illustrates an extreme case for channel K, suggesting some over-estimation of the flux in a complex background scenario with FIT-C and FIT-L, but not with SPSR, as reported in Figure~\ref{fig:anomaly} in the simulation.

\section{\label{sec:repres}Reproducible research}

\begin{table*}[htbp]
  \centering
  \begin{tabular}{@{} ll @{}} 
  Product Name & Description\\
  \hline
  \emph{WMAP} 9-year  products:\\
  WMAP9\_data\_chanK\_ptsub.fits   & SPSR point-source-free  \emph{WMAP} 9-year K channel\\
  WMAP9\_data\_chanKa\_ptsub.fits   & SPSR point-source-free \emph{WMAP} 9-year Ka channel\\
  WMAP9\_data\_chanQ\_ptsub.fits    & SPSR point-source-free \emph{WMAP} 9-year Q  channel \\
  WMAP9\_data\_chanV\_ptsub.fits    & SPSR point-source-free \emph{WMAP} 9-year V  channel \\
  WMAP9\_data\_chanW\_ptsub.fits    & SPSR point-source-free \emph{WMAP} 9-year W  channel \\
    \hline 
  Software products (IDL):\\
  {\tt mrs\_sparse\_pointsource\_removal.pro}   & removes point sources from a spherical map\\  & (requires {\tt HealPix} and {\tt iSAP}). \\
  {\tt WMAP9\_DATA\_routines.pro}   & routines to obtain the \emph{WMAP} 9-year maps \\  & and beams required by SPSR (requires {\tt HealPix} and {\tt iSAP}).\\
  {\tt wmap9\_remove\_point\_sources.pro}  & script that applies SPSR to all \emph{WMAP} 9-year channels\\  & (requires {\tt HealPix} and {\tt iSAP}).\\
  \hline

  \end{tabular}
  \caption{List of products made available in this paper in the spirit of reproducible research, available at \url{http://www.cosmostat.org/Products}.}
  \label{tab:reproducible_ps}
\end{table*}

In the spirit of participating in reproducible research, we make all codes and resulting products that constitute the main results of this paper public. In Table \ref{tab:reproducible_ps} we list all products that are made freely available as a result of this paper, and which are available at \url{http://www.cosmostat.org/Products}.

For the five \emph{WMAP} 9-year channels, SPSR was  applied using the following command line in the open source package {\tt iSAP} software\footnote{\url{http://jstarck.free.fr/isap.html}}:\\
{\tt\bf
\noindent $>$  map\_psfree =\\
 \indent mrs\_sparse\_pointsource\_removal(Map, GalMask, BeamInfo,StdMap,  Niter=Niter).} 

For Ka to W maps, Niter was fixed to 9750, and to 13350 for the K channel. Galmask is the \emph{WMAP} point-source catalogue mask\footnote{\url{http://lambda.gsfc.nasa.gov/data/map/dr5/ancillary/masks/wmap_point_source_catalog_mask_r9_9yr_v5.fits}}, StdMap is the map containing the standard deviation of the noise per pixel, and BeamInfo is a structure containing the \emph{WMAP} 9-year beam at each point-source position. The code {\tt WMAP9\_DATA\_routines.pro} can be used to obtain the required maps and beams and to call the {\tt mrs\_sparse\_pointsource\_removal} routine to derive the final products, as illustrated in the script {\tt wmap9\_remove\_point\_sources.pro}.

\section{Conclusions}
\label{sec:ccl}

We proposed a new approach for detected point-source flux estimation and subtraction. Compared with the standard approach, which estimates point-source flux according to local low-order polynomial models of the background, the proposed technique is based on a global modelling of the background, which is assumed to be sparse in spherical harmonics to better capture its fluctuations. Bright point-source emissions decrease the sparsity of this background, which is the key phenomenon driving the separation process.
An algorithm was adapted from recent convex optimization developments to solve the corresponding inverse problem.

We evaluated the proposed technique as well as techniques used in \emph{WMAP} and \emph{Planck} collaboration on realistic simulations of the \emph{WMAP} microwave sky. In a noise-limited catalogue, except for channel W, where noise leads to faint differences between the estimates and a slightly poorer estimate for the proposed approach for low-flux sources (about $6\%$), our approach out-performs local polynomial fitting. In the internally derived catalogue, SPSR also consistently leads to
\begin{itemize}
\item[-] the lowest biases in the first three channels (up to $100\textrm{mJy}$ lower bias for sources $< 1\textrm{Jy}$ in channel K)
\item[-] the lowest RMSE in all channels  and all flux bands (with values decreased by at least $5\%$, $25\%$, $28\%$,$7\%$,
and $3\%$ and up to $35\%$, $28\%$,  $32\%$, $9\%,$ and $14\%$ for channels K to W) 
\item[-] more robust point-source subtraction in complex background, as illustrated in specific examples
\item[-] consequently better multichannel estimates of the point-source fluxes, which is useful to estimate more accurate spectra. 
\end{itemize}

This technique was finally applied to the \emph{WMAP} 9-year data deconvolved with the beam asymmetries and the resulting point-source-subtracted maps are available on-line. We focused in this paper on point sources. Even if the proposed model for extended compact sources is rather simple in SPSR, it is sufficient to derive a background accurate enough to improve on point-source flux recovery compared with FIT-C  or  FIT-L. However, better modelling of the extended compact sources is still required if they need be subtracted from the data. We left this task for future work.

\section*{Acknowledgment}
This work was funded by the European Research Council grant SparseAstro (ERC-228261), and the Swiss National Science Foundation (SNSF).
We used the  Healpix\footnote{\url{http://healpix.sourceforge.net}} software \citep{Gorski05}, the iSAP\textcolor{red}{\footnotemark[1]} software, and \emph{WMAP} data\footnote{\url{http://map.gsfc.nasa.gov}}.

\bibliographystyle{aa}

\end{document}